\documentclass[journal=jacsat,manuscript=article,maxauthors=7,
etalmode=truncate]{achemso}
\setkeys{acs}{maxauthors=7}
\setkeys{acs}{etalmode=truncate}

\usepackage{amsmath}
\usepackage{amsmath,amssymb}
\usepackage{graphicx}
\usepackage{caption}
\usepackage{color}
\usepackage{dcolumn}
\usepackage{bm}
\usepackage{float}
\usepackage{subfig}
\usepackage{soul}
\usepackage[normalem]{ulem}
\usepackage{microtype}
\usepackage[version=3]{mhchem} 
\usepackage{booktabs}
\usepackage{multirow}
\usepackage{booktabs}
\usepackage{longtable}
\usepackage[table]{xcolor} 

\usepackage{xr}

\makeatletter
\newcommand*{\addFileDependency}[1]{%
  \typeout{(#1)}%
  \@addtofilelist{#1}%
  \IfFileExists{#1}{}{\typeout{No file #1.}}%
}
\makeatother

\newcommand*{\myexternaldocument}[1]{%
  \externaldocument{#1}%
  \addFileDependency{#1.tex}%
  \addFileDependency{#1.aux}%
}

\myexternaldocument{si}

\usepackage{rotating}

\SectionNumbersOn
\usepackage[dvipsnames]{xcolor}

\author{Eric D. Boittier} \affiliation[University of Basel]{Department
  of Chemistry, University of Basel, Klingelbergstrasse 80, CH-4056
Basel, Switzerland.}

\author{Markus Meuwly} \affiliation[University of Basel]{Department of
  Chemistry, University of Basel, Klingelbergstrasse 80, CH-4056
Basel, Switzerland.}  \email{m.meuwly@unibas.ch}

\title{Efficient, Equivariant Predictions of Distributed Charge Models}

\begin{document}

\begin{tocentry}

  \includegraphics[width=6.7cm]{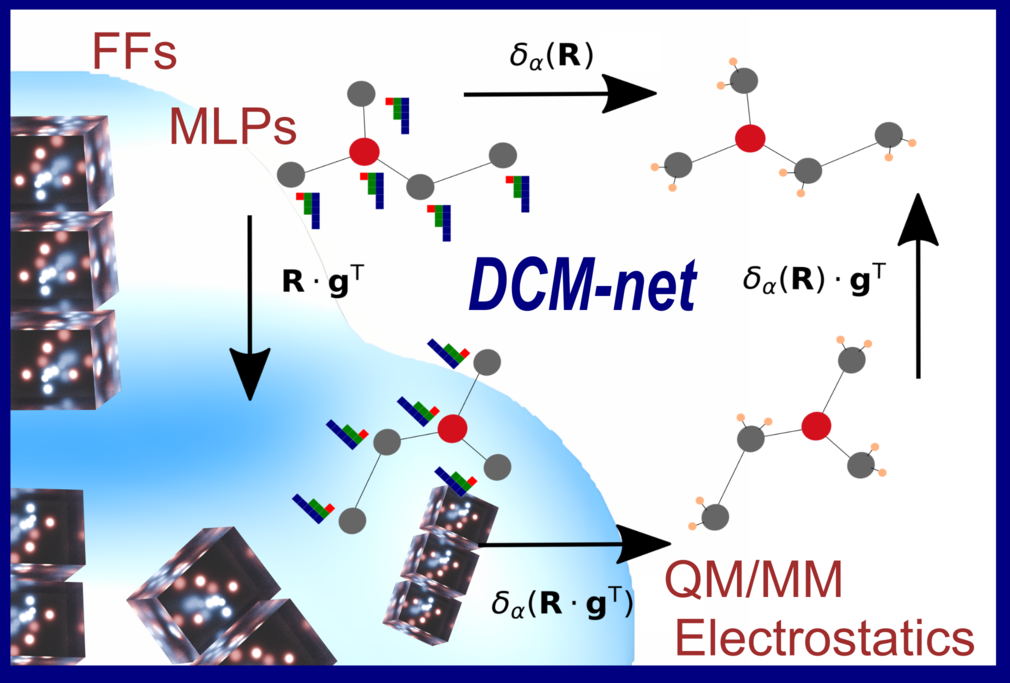}

\end{tocentry}

\begin{abstract}
A machine learning (ML) based, equivariant neural network for constructing distributed charge models (DCMs) of arbitrary resolution - DCM-net - is presented. DCMs efficiently and accurately model the anisotropy of the molecular electrostatic potential (ESP) and go beyond the point charge representation used in conventional molecular mechanics (MM) energy functions. This is particularly relevant for capturing the conformational dependence of the ESP (internal polarization) and chemically relevant features such as lone pairs or $\sigma-$holes. Across conformational space, the
learned charge positions from DCM-net are stable and continuous. Across the QM9 chemical space,
two-charge-per-atom models achieve accuracies comparable to fitted
atomic dipoles for previously unseen molecules (0.75 (kcal/mol)/$e$). Three- and
four-charge-per-atom models reach accuracies competitive with atomistic
multipole expansions up to quadrupole level (0.55 (kcal/mol)/$e$). Pronounced improvements of the ESP are found around O- and F-atoms both of which are known to feature strongly anisotropic fields, and for aromatic systems. Across the QM9 reference data set, molecular dipole moments improve by $\sim 0.1$ D compared with fitted monopoles.  Transfer learning on dipeptides yields a 0.2 (kcal/mol)/$e$ ESP improvement for unseen samples and a
two-fold MAE reduction for molecular dipole versus fitted monopoles. Overall, DCM-net offers a fast and physically meaningful approach to generating
distributed charge models for running pure ML- or mixed ML/MM-based molecular simulations.

\end{abstract}
\date{\today}

\section{Introduction}
The electrostatic potential (ESP) around a molecule can be used to explain
a wide range of molecular properties related to structure, bonding,
stability, and reactivity, among
others.\cite{Bayly:1993,stone:2013,murray:2011,politzer:2002,suresh:2022}
The ESP can be highly anisotropic leading to a variety of chemically
relevant features including "lone pairs" and "$\sigma-$holes", for
example, which cannot be modeled with atom-centered
charges.\cite{Ibrahim:2011,schyman:2012,Yan:2017,Kolar:2012} By
reducing the representation of the local electrostatic environment to
atomic multipoles or point charges (PCs), the true anisotropic
electrostatic potential of the molecule is
neglected.\cite{Kramer:2014} A successful approach to modeling the
molecular electrostatic potential has been to use atom-centered
multipoles (MTPs) to a given order.\cite{Kramer:2012,Bereau:2013a,Bereau:2013b,Kramer:2014} Often, this expansion is truncated
at the atomic quadrupole moment for all atoms except hydrogens for
which a PC representation is usually
sufficient.\cite{price:1990,MM.mbco:2008,ponder:2010,Kramer:2014}
Including higher order atomic multipoles has been successful in
crystal structure prediction\cite{Day:2005} and modeling interaction
energies using force
fields.\cite{Devereux:2014a,dykstra:1993,heindel:2025} These atomic
multipole expansions can be obtained from the electron density of the
wavefunction, using techniques such as the Minimal Basis Iterative
Stockholder (MBIS) procedure\cite{Verstraelen:2016}.\\

\noindent
The quantum theory of atoms in molecules (QTAIM) suggests features in
the molecular density, such as bond critical points, which are well
described using distributed multipoles.\cite{Misquitta:2014} Likewise,
PC-based energy functions have been supplemented by off-center PCs in
the OPLS force field to model the anisotropy of the
ESP\cite{schyman:2012,Yan:2017}, sometimes referred to as `extra
charge' or lone pair sites, introduced in the AMBER force field to
better describe hydrogen bonding and halogen
interactions\cite{Weiner:1984,Dixon:1997}, or virtual interaction
sites in GROMACS.\cite{gromacs:2024} More generally, an atomic
multipole expansion can be approximated using multiple point charges arranged around an atomic center, which leads to "distributed charge models" (DCMs).\cite{Devereux:2014b} Formally, to obtain a
DCM-representation of the multipole expansion up to the quadrupole level,
six charges per atom are required.\cite{Devereux:2014b} Like atomic
multipoles,\cite{Kramer:2012} improvements in interaction energies have also been
observed with distributed charge models.\cite{Devereux:2024} The
derived multipoles are sensitive to the local atomic environment and
therefore exhibit pronounced conformational dependence\cite{Kramer:2012,
  stone:1995} suggesting that a static representation of atomic
multipoles may not be sufficient in some cases.  Geometry-dependent distributed charge models to capture effects of bond polarization have also been
considered.\cite{Boittier:2022, Boittier:2024} When modeling the
molecular ESP, significant redundancy in electrostatic parameters has
been observed\cite{Jakobsen:2014}.  This has justified the use of
minimal distributed charges \cite{MM.dcm:2017} which involved fitting
charges to atomic environments and combining these initial guesses
using differential evolution (DE) to obtain a model with low
complexity and sufficient quality.\cite{MM.dcm:2017}\\

\noindent
In force-field development, it is often desirable to obtain
`transferable' parameters across some predefined chemical space
whereby topologically similar atoms are assigned the same `atom type'
to reduce the complexity of the model.\cite{Weiner:1984, ponder:2010,
  Warshel:2007, Devereux:2024} Ideally, these shared parameters should
be accurate for molecules not explicitly included during
parametrization.\cite{Bereau:2013b, Bereau:2015} Transferable
multipoles were investigated by Bereau and
coworkers\cite{Bereau:2013b, Bereau:2015} using predefined reference
frames based on handcrafted atom types\cite{Kramer:2012,Bereau:2013b}
Alternatively, transferability can be obtained without predefined
reference frames by using a molecular frame of reference, which can be
derived from the principal moments of inertia.\cite{Zare:1988, BrinkSatchler} Neural networks with
directional features can learn a `representation' of local reference
frames from the data which may improve
transferability.\cite{Thrlemann:2022} The difference between
`hand-crafted' versus learnt local reference frames is analogous to
manually constructed features used in Behler--Parrinello atomistic
neural networks versus data-driven features learnt by message-passing
networks such as PhysNet\cite{Unke:2019}.\\

\noindent
Geometric deep learning\cite{Bronstein:2017, Duval:2024} has emerged
as a promising strategy for problems where strong inductive priors
stemming from the laws of physics, such as the invariance of energy
with respect to rotations and translations, can be leveraged to create
efficient, performant, and physically meaningful
models.\cite{Bronstein:2017, Duval:2024} Equivariant neural networks
are data-efficient, universal approximators for tasks where inherent
symmetries exist, such as rotational equivariance.\cite{Schtt:2021,
  Batatia:2022, Brandstetter:2022} For example, vector quantities such
as dipole moments should rotate with the molecule.\cite{Zare:1988,
  BrinkSatchler} Equivariant neural networks have been applied to the
prediction of atomic multipole expansions.\cite{Thrlemann:2022} By
including higher order many-body messages based on spherical
harmonics, techniques such as Multiple Atomic Cluster Expansion
(MACE)\cite{Batatia:2023}, as well as equivariant attention
mechanisms\cite{Frank:Unke}, have shown state-of-the-art accuracy in a
variety of tasks, such as predicting molecular dipole moments.  This
approach has also been generalized to Cartesian tensors, which may
require fewer parameters.\cite{Simeon:2023}\\

\noindent
Kernel-based methods using
rotationally equivariant features have also been used to model the
molecular dipole moment.\cite{Sun:2022} If only scalar properties are
of interest, traditional deep neural networks are sufficient, e.g.,
for predicting electrostatic parameters for the polarizable Drude
classical force field.\cite{Kumar:2022} Although strictly equivariant
models have garnered significant attention, alternatives which weaken
these requirements have been found to provide better performance when
large amounts of training data and compute resources are
available.\cite{Pozdnyakov:2023, Wang:2024} For instance, the Point
Edge Transformer,\cite{Pozdnyakov:2023} a graph neural network where
each message-passing layer is given by an arbitrarily deep
transformer.  Furthermore, strategies such as data
augmentation\cite{Glick:2021} have been employed whereby the input and
output features are rotated arbitrarily during
training.\cite{Glick:2021} Learning a representation of the density
has been explored via construction of the Hamiltonian matrix in a
block-wise manner from equivariant representations\cite{Unke:2021}, as
well as grid based electron density learning schemes\cite{Koker:2024,
  Li:2024}.\\

\noindent
The present work describes ``DCM-net'' (distributed charge model
network) which is an equivariant neural network for the prediction of
distributed charges - trained to minimize the error of the
molecular ESP. 
The model predicts up to $n_{\mathrm{DC}}$ distributed charges per
atom.  The performance of the model at reproducing the ESP and
molecular dipole moment is assessed on a hold-out set of unseen
structures to assess the model's transferability.

\section{Methods}
First, a cursory background on molecular electrostatics is
presented, then the SO(3)-equivariant architecture is described, along
with data curation, training objectives, and various evaluations.

\subsection{Electrostatics, Multipoles, and Distributed Charges}
\begin{figure}[!ht]
  \centering
  \includegraphics[width=0.5\linewidth]{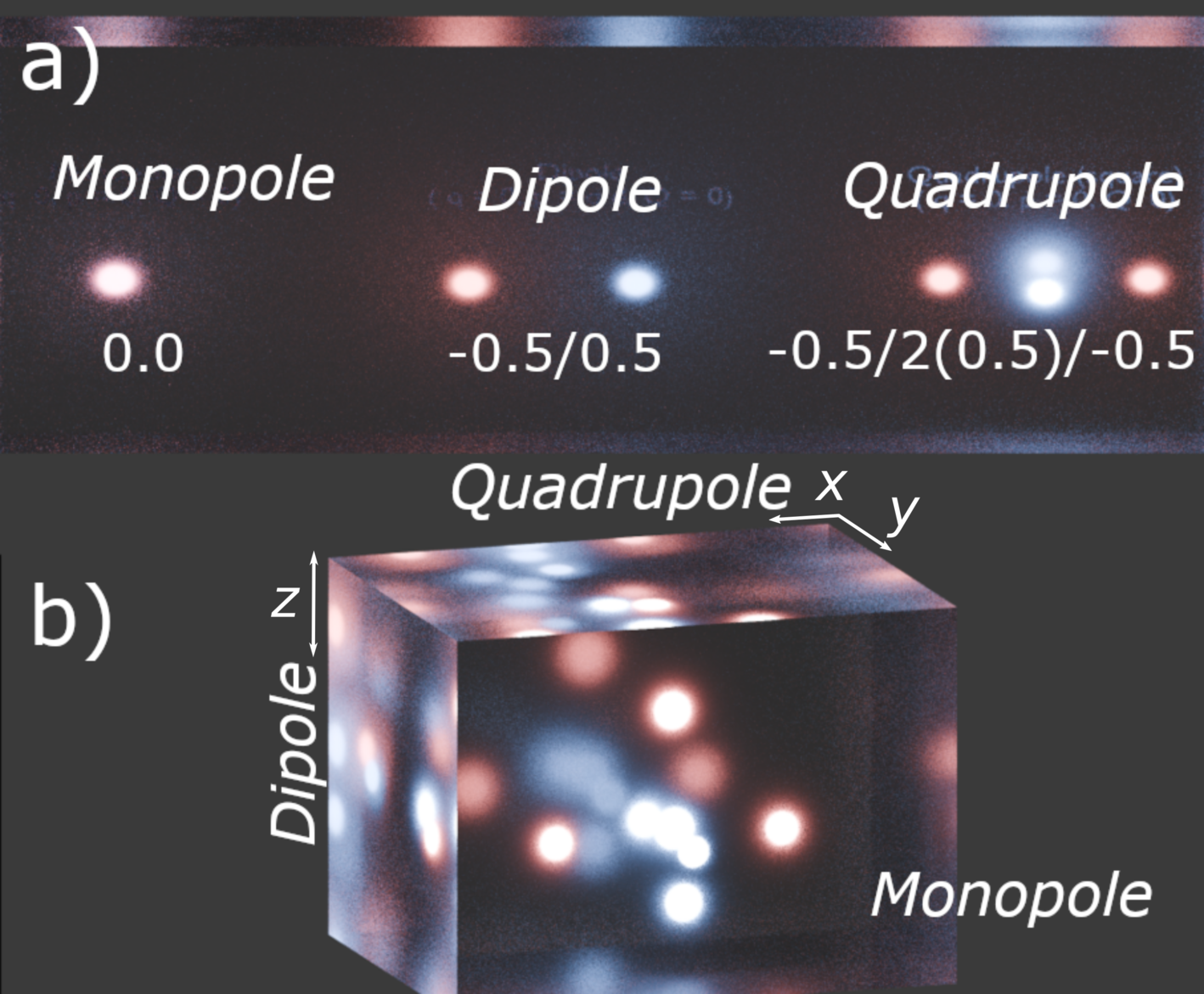}
  \caption{Reproducing the ESP: a graphical illustration of equations
    \ref{eq:mono}, \ref{eq:dip}, and \ref{eq:quad} using a computer
    render of colored `point lights' in glass. a) The standard point
    charge representations of the monopole, dipole, and quadrupole
    moments. b) Aligning the dipole perpendicularly to the quadrupole
    directions gives a distributed charge solution for a system with
    vanishing monopole moment. Distributed charges offer a parameter
    efficient approach to fitting the ESP in volumes of interest
    (outside the glass, i.e. van der Waals region).}
  \label{fig:eg}
\end{figure}

The electrostatic potential (ESP) at a point $\mathbf{r}$ arising from
a continuous charge distribution $\rho(\mathbf{r}')$, omitting the
prefactor $(4\pi \varepsilon_0)^{-1}$ for clarity, is
\begin{equation}
  \text{ESP}(\mathbf{r}) \sim \int_{V'}
  \frac{\rho(\mathbf{r}')}{|\mathbf{r} - \mathbf{r}'|} dV'
\end{equation}
Expanding the ESP in a Taylor series about a chosen center, typically
the center of charge or an atomic nucleus, yields the familiar
multipole expansion.\cite{jackson} In spherical coordinates, and
truncated at maximum angular momentum $\ell_{\text{max}}$, the ESP at
coordinates $(r, \theta, \phi)$ can be approximated by
\begin{equation}
  \text{ESP}(r, \theta, \phi) \approx
  \sum_{\ell=0}^{\ell_{\text{max}}} \sum_{m=-\ell}^{\ell}
  Q_{\ell}^{m} Y_{\ell}^{m}(\theta, \phi) R_{\ell}(r)
\end{equation}
Here, $Y_{\ell}^{m}(\theta, \phi)$ are spherical harmonics,
$R_\ell(r)$ are radial functions determined by the order of the
multipole, and the coefficients $Q_{\ell}^{m}$, the multipole moments,
encode the angular components of the charge
distribution.\cite{jackson, stone:2013, buckingham:1968} These
coefficients can be derived from $\rho(\mathbf{r}')$, though they
depend on the choice of partitioning scheme when applied to atoms in
molecules, and are therefore not unique.\cite{stone:2013, stone:1995}
The lowest-order terms correspond to familiar physical quantities (see
Figure \ref{fig:eg}a; $\ell = 0$: monopole (total charge), $\ell = 1$:
dipole, $\ell = 2$: quadrupole).\\

\noindent
A finite set of symmetrically arranged monopoles can exactly reproduce
the low-order multipoles; for example, a system of six judiciously
placed charges would be correct up to quadrupole order, defining the
monopole (placed at the center of mass), dipole (two charges placed at
some coordinates scaled by $r$), and selected quadrupole components
(placed at some coordinate scaled by $r^2$).\cite{Devereux:2014b,
  MM.dcm:2017} In this fashion, the number of charges would scale as
$O((1+\ell)^{2} - l+1)$.  For example, the total charge (referred to
here as the `monopole') is
\begin{equation}
  Q_0^0 = q = \sum_{i=1}^{6} q_i
  \label{eq:mono}
\end{equation}
The dipole moments in the $z$, $x$, and $y$ directions (expressed
using real spherical harmonic convention for $Q_1^m$) are
\begin{equation}
  Q_1^0 = \mu_z = \sum_{i=1}^{6} q_i r_{z,i}, \quad
  Q_1^{-1} = \mu_x = \sum_{i=1}^{6} q_i r_{x,i}, \quad
  Q_1^{1} = \mu_y = \sum_{i=1}^{6} q_i r_{y,i}
  \label{eq:dip}
\end{equation}
and the (non-redundant) quadrupole terms are
\begin{equation}
  Q_2^0 = \Theta_{zz} = \sum_{i=1}^{6} \frac{1}{2} q_i \left(
  3r_{z,i}^2 - r_i^2 \right), \quad
  Q_2^{2c} = \frac{1}{\sqrt{3}} (\Theta_{xx} - \Theta_{yy}) =
  \sum_{i=1}^{6} \frac{3}{4} q_i \left( r_{x,i}^2 - r_{y,i}^2 \right)
  \label{eq:quad}
\end{equation}
Here, $r_{x,i}, r_{y,i}, r_{z,i}$ are the Cartesian coordinates of
each point charge $q_i$ in a local coordinate frame (typically the
principal axis system), and $r_i^2 = r_{x,i}^2 + r_{y,i}^2 +
r_{z,i}^2$.  This construction is illustrated in Figure~\ref{fig:eg}
using a computer render of colored `point lights' in glass.  The first
three moments are reproduced when the dipole moment is aligned along the
$z-$direction, orthogonal to the quadrupole moment in the $x-$ and $y-$directions.  While this example uses six charges to exactly reproduce
the leading multipole moments, fewer charges can be used to generate
approximate representations of the ESP, giving rise to the minimal
distributed charge model (MDCM).\cite{MM.dcm:2017} The use of
distributed point charges in this way underlies the DCM formalism
employed in the present work.\\

\noindent
This motivates a compact representation of anisotropy via a small set
of off-center charges; DCM-net learns their positions and magnitudes
from local environments while preserving rotational
behavior.\cite{schyman:2012, Yan:2017, Dixon:1997}\\

\begin{figure}[!ht]
  \centering
  \includegraphics[width=0.623\linewidth]{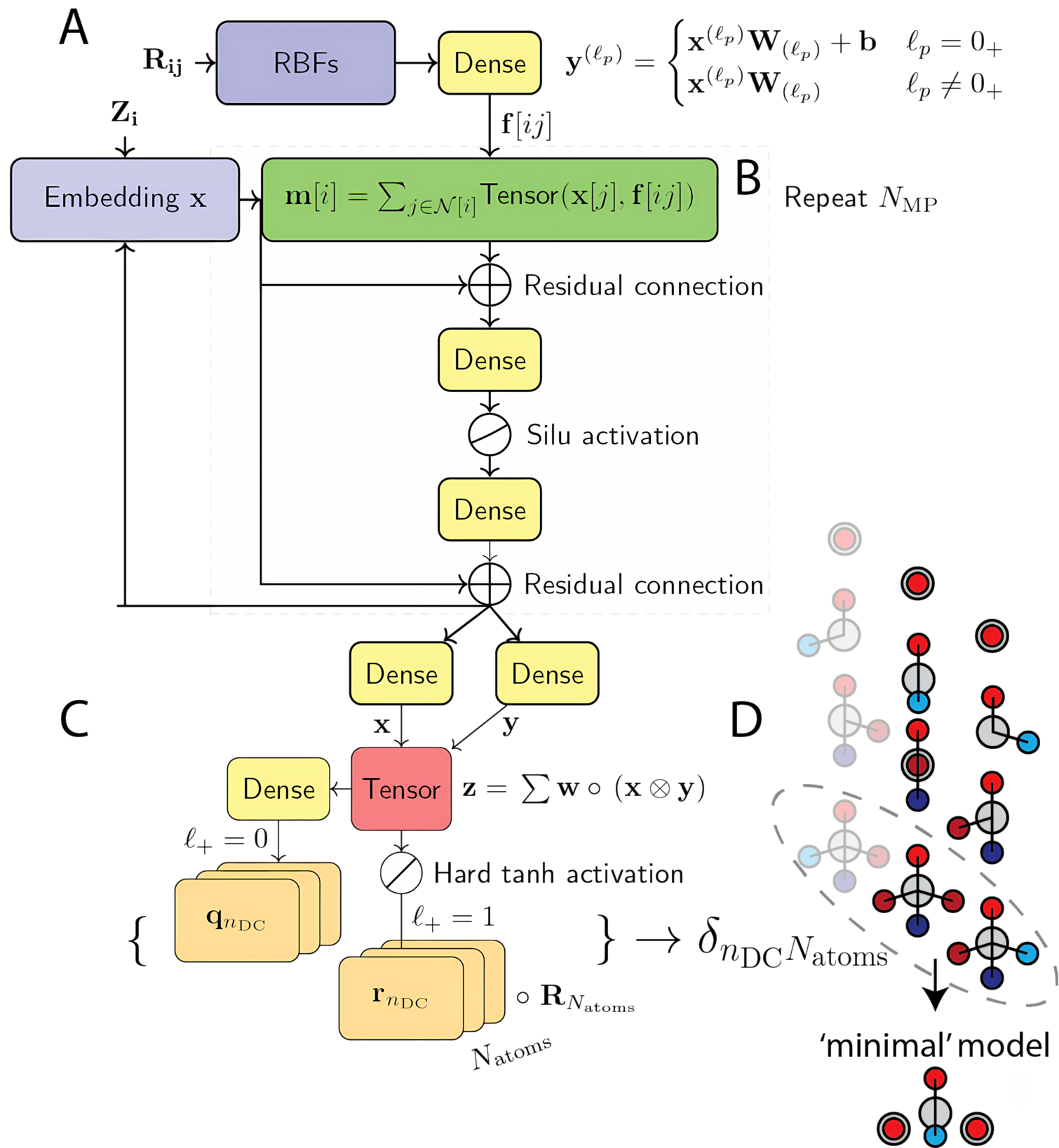}
  \caption{The architecture of DCM-net model. (A) The inputs to the
    network are atomic numbers and positions expanded into atom-atom
    distances. (B) During the message passing phase, the hidden
    representation is updated over $N_{\rm MP}$ iterations using the
    message passing operation (green).
    Dense layers (Eq. \ref{eq:dense}) and tensor products (Eq.
    \ref{eq:tensor}) are shown in yellow and red, respectively. (C)
    The final output is split between scalar features (monopoles) and
    vector-like features (charge displacements relative to atomic centers).
    (D) Predictions using $n_{\rm dc}$ distributed
    charges (red, blue) per atom (gray) can be combined and optimized
    to create 'minimal' models with acceptable accuracy and improved
  computational efficiency, and can be used in molecular dynamics simulations.}
  \label{fig:arch}
\end{figure}

\subsection{SO(3)-Equivariant Graph Neural Network}
DCM-net predicts the charge and position of a set of distributed
monopoles to represent the molecular electrostatic potential (ESP).
It is implemented in JAX\cite{jax} using the e3x
package\cite{Unke:2024}. The model employs a message-passing
architecture composed of equivariant dense layers and tensor product
operations, described below and illustrated in Figure~\ref{fig:arch}.
The choice of an equivariant architecture is motivated, in the SI (Figures \ref{sifig:example} and \ref{sifig:hiddenstates}), by
way of a toy example.\\

\noindent
An equivariant dense layer operates analogously to a standard
multilayer perceptron when propagating scalar (rotation-invariant)
features, with weights $\mathbf{W}$ and biases $\mathbf{b}$. For
tensor-valued features (e.g. the dipole moment vector $\mathbf{D}$),
the bias term is omitted to preserve SO(3) equivariance (see `Dense
layer' in Figure~\ref{fig:arch}A).\cite{Schtt:2021, Batatia:2022} Let
$\ell$ denote the degree of the spherical harmonic (or equivalently,
the rank of the irreducible representation), and let $p = (-1)^\ell$
denote the parity under inversion. The layer transformation is:

\begin{equation}
  \mathbf{y}^{(\ell)} =
  \begin{cases}
    \mathbf{x}^{(\ell)}\mathbf{W}^{(\ell)} + \mathbf{b}, & \ell = 0
    \ (\text{scalar}) \\
    \mathbf{x}^{(\ell)}\mathbf{W}^{(\ell)}, & \ell > 0 \ (\text{tensor})
  \end{cases}
  \label{eq:dense}
\end{equation}

\medskip
\noindent
Tensor features are combined via tensor product operations (denoted
``Tensor" in Figure~\ref{fig:arch}B and C), which mix compatible input
irreducible representations based on angular momentum coupling rules (see below).\cite{Zare:1988}
The tensor product uses learnable weights $\mathbf{w}_{(l_1, l_2,
  l_3)}$, and computes the coupled output features
$\mathbf{z}^{(l_3)}$:

\begin{equation}
  \mathbf{z}^{(l_3)} = \sum_{(l_1, l_2)\in V}
  \mathbf{w}_{(l_1, l_2, l_3)} \circ \left(
    \mathbf{x}^{(l_1)} \otimes^{(l_3)} \mathbf{y}^{(l_2)}
  \right)
  \label{eq:tensor}
\end{equation}

\noindent
The tensor product $\otimes^{(l_3)}$ couples input features using
Clebsch–Gordan coefficients\cite{Zare:1988,BrinkSatchler}:
\begin{equation}
  \left( \mathbf{x}^{(l_1)} \otimes \mathbf{y}^{(l_2)}
  \right){m_3}^{(l_3)} = \sum{m_1=-l_1}^{l_1} \sum_{m_2=-l_2}^{l_2}
  C_{l_1, l_2, l_3}^{m_1, m_2, m_3} , x_{m_1}^{(l_1)} y_{m_2}^{(l_2)}
  \label{eq:coupling}
\end{equation}
Here, $C_{l_1, l_2, l_3}^{m_1, m_2, m_3}$ are the Clebsch–Gordan
coefficients that couple angular momenta $l_1$ and $l_2$ to $l_3$ and similarly for the projections $m_1$, $m_2$, and $m_3$, which amounts to ensure that the coupled feature transforms according to
the $l_3$ irreducible representation of SO(3).\\

\noindent
In Eq. \ref{eq:coupling} the set $V$ contains combinations of $(l_1, l_2)$ such that $C_{l_1, l_2, l_3}^{m_1, m_2, m_3} \neq 0$
given output $l_3$. Each
$\mathbf{w}_{(l_1, l_2, l_3)} \in \mathbb{R}^{1 \times 1 \times F}$ is
a learnable weight shared across features (broadcast over all
atoms). The element-wise product $\circ$ denotes broadcasting over
tensor dimensions.\\

\noindent
Input features $\mathbf{x} \in \mathbb{R}^{P_1 \times (L_1+1)^2 \times
  F}$ and $\mathbf{y} \in \mathbb{R}^{P_2 \times (L_2+1)^2 \times F}$
are indexed over irreducible representation channels $(L_i + 1)^2$
corresponding to the number of real spherical harmonic components up
to and including angular momentum $L_i$. The parity index $P_i =
2$ if pseudotensors are included, and $P_i = 1$ otherwise. The output
is $\mathbf{z} \in \mathbb{R}^{P_3 \times (L_3+1)^2 \times F}$.\\

\noindent
Initial atomic embeddings $\mathbf{x}$ are obtained from atomic
numbers $\mathbf{Z}$ via a learnable embedding layer. Interatomic
displacement vectors $\mathbf{R}_{ij}$ are computed from atom
positions and expanded in a product basis consisting of radial and
angular terms:
\begin{equation}
  \Phi_{lm}(r, \theta, \phi) = B_n(r) Y_{lm}(\theta, \phi)
\end{equation}
where $Y_{lm}$ are real spherical harmonics and $B_n(r)$ are radial
basis functions. Following previous work,\cite{Unke:2024} Bernstein polynomials for
$B_n(r)$ are used, due to their compact support and flexibility when modeling exponential decay as a basis with spherical harmonics parametrized by
the maximum degree and the number of basis functions. Reciprocal
Bernstein radial functions modified by a smooth cutoff function were
also included in the features.\cite{Unke:2024} The standard message
passing scheme involves the application of the graph convolution
$\phi_e$ on edge features $m_{ij}$, which are summed into node
features $m_i$ used to update the hidden embedding $h_i$:
\begin{align}
  m_{ij} &= \phi_e(h_i^l, h_j^l, a_{ij}) \\
  m_i &= \sum_{j \in \mathcal{N}(i)} m_{ij} \\
  h_i^{l+1} &= \phi_h(h_i^l, m_i)
  \label{eq:message}
\end{align}
Each message passing step (Eq. \ref{eq:message}) is followed by a
dense layer, a SiLU activation function, and another dense layer,
which is added to the embedding and is repeated for $n_{\mathrm{MP}}$
steps.\\

\noindent
Following the iterative message passing, the atom-wise features
undergo further refinement where the output of two dense layers are
combined using their tensor product (Eq. \ref{eq:tensor}). The
transformation of features to the desired output involves a dense projection which passes the hidden state to a feed-forward neural network with a feature size of $n_{\mathrm{DC}}$ per atom. The
\texttt{hardtanh} activation function scaled by a constant was used to
limit the displacements of the distributed charges. The scaling factor
used was 0.175 \AA~ to limit displacements to no more than 0.3
\AA~away from the atomic centers although this hyperparameter may be
adjusted.\\

\noindent
The displacements of the distributed charges are obtained from the
$\ell_p=1_-$ features, while the $\ell_p=0_+$ (scalar) features are
input into a final dense layer (with an element specific bias) before
the final readout step where $n_{\mathrm{DC}}$ distributed monopoles
for each atom are obtained. Here the subscript denotes the parity of
the feature, indicating the sign of the feature under inversion (e.g.,
$\ell_p=1_-$ denotes the odd parity features which change sign under
inversion). The final position of each distributed charge,
$\boldsymbol{\delta}$, is the sum of the atom's position $\mathbf{R}$
and the displacement vector obtained in the final readout
$\mathbf{r}$.\\

\noindent
Aside from rotational equivariance, the conservation of charge in an isolated system is another physical law that needs to be captured by the model.  The correct total
charge of the molecule, $Z$, may be fixed by subtracting the mean
predicted charge per $n_{\rm DC}$ charges from the average charge charge per atom and adding the average remaining charge  $\langle q \rangle$ to each of the charges
\begin{equation}
  \langle q \rangle = \frac{Z}{N_{\mathrm{atoms}}} - \frac{1}{
  n_{\mathrm{DC}}} \sum_{i}^{n_{\mathrm{DC}}} q_i
  \label{eq:chgeq}
\end{equation}

This transformation is applied after training, and in the read out of
the validation loss. Empirically,\cite{Unke:2021, Schtt:2021,
  Thrlemann:2022} including this constraint in the loss function has
been shown to hinder training by slowing convergence and leading to models with larger errors. Penalizing the
total molecular charge is more stable and to be preferred. The
regularization of Eq. \ref{eq:chgeq} can be applied during inference
if required.\\

\noindent
The loss function includes terms
relating the root-mean-squared error of the electrostatic potential
evaluated on the grid, $\mathrm{RMSE_{ESP}}$, between the reference
molecular ESP$_{\rm ref.}$ obtained from quantum chemical calculations
and the ESP generated by the distributed charge model.
The distributed charges are described by
$\boldsymbol{\delta}$ (a
vector consisting of a position and charge) evaluated over $N_{\rm
  grid}$ grid points $\mathbf{y}_i$, where the model ESP is calculated
as the sum of the Coulomb interactions between each (distributed)
charge and a probe charge of 1$e$.
\begin{equation}
  \mathrm{RMSE_{ESP}}(\boldsymbol{\delta}) := \sqrt{ \frac{1}{N_{\rm
    grid}} \sum^{N_{\rm grid}}_{i=1} \left [ {\rm ESP}_{\rm
      ref.}(\mathbf{y}_i) - {\rm ESP}_{\rm
      model}(\mathbf{y}_i, \boldsymbol{\delta})
  \right ] ^{2}}
\end{equation}

\noindent
An additional (optional) term in the loss function corrects the
molecular dipole moment generated by the distributed charges
($\mathbf{D}_{\mathrm{DC}} = \sum_{i=0}^{N_{\mathrm{atoms}}
  n_{\mathrm{DC}}} q_{\mathrm{DC},i} (\mathbf{R}_{\mathrm{CoM}} -
\mathbf{\delta}_{\mathrm{DC},i})$), weighted by $w_{\mathrm{D}}$ is
included in the loss function (Eq. \ref{eq:loss1}):
\begin{equation}
  \mathcal{L} = w_{\mathrm{ESP}} \cdot \mathrm{RMSE_{ESP}} +
  w_{\mathrm{q}} \cdot
  \frac{1}{N_{\mathrm{atoms}}}\sum^{N_{\mathrm{atoms}}}_{i=0}
  \bigg(M^0_{\mathrm{ref,i}} - \sum_{j=0}^{n_{\mathrm{DC}}}
  M^0_{i,j}\bigg)^2 + w_{\mathrm{D}} \cdot \frac{1}{3}
  \sum_{\alpha = \{x,y,z\}} ({D}_{\mathrm{ref,\alpha}} -
  {D}_{\mathrm{DC},\alpha})^2
  \label{eq:loss1}
\end{equation}
The mean absolute error between the MBIS monopoles and the per-atom
sum of the distributed charges, weighted by $w_{\mathrm{q}}$ is also
included. This term ensures that the distributed charges are able to
reproduce the MBIS monopoles, however, any baseline charge assignment
scheme can be used.  Regularizing the sum of atomic charges to match
the MBIS monopoles is an additional constraint also used in methods
such as RESP\cite{Singh:1984, Bayly:1993}, which uses an L$_2$ penalty
to enforce consistent atomic charges even on buried atoms (i.e. no
accessible volume or grid points associated with said atom), as ESP
fitting is ill-posed in this case.\\

\noindent
For vectorial quantities, such as molecular dipole vectors
$\mathbf{D}$ and forces $\mathbf{F}$, the average per-component RMSE
and MAE were used and reported in the loss functions and results,
respectively.
The MBIS molecular dipole is calculated as $\mathbf{D}_{\mathrm{MBIS}} = \sum_{i=0}^{N_{\mathrm{atoms}}
  } q_{i} (\mathbf{R}_{\mathrm{CoM}} -
\mathbf{R}_{i}) + \sum_{i=0}^{N_{\mathrm{atoms}}} \bf{M}^{1}_i$ where $\bf{M}^{1}_i$ is the atom-centered dipole moment on atom $i$ obtained from the MBIS multipole expansion.
\\

\subsection{Data Generation}
Training data compositions split tests into two main streams: (1)
\textit{conformational space}, where model a parametrizes electrostatics
for many conformations of a single molecule, and (2) \textit{chemical
  space}, where the model learns a shared electrostatic representation
for a variety of molecules.\\

\noindent
For models covering \textit{conformational space}, 10,000 conformers
were generated for carbon dioxide enumerating 100$^{3}$ combinations
of the two CO bond lengths ($1.0 < r < 1.80$ \AA) and OCO angle
($120^{\circ} < \theta < 180^{\circ}$) in a Z-matrix representation.
Calculations were performed in Molpro\cite{werner.molpro:2020} at the
MP2/aug-cc-pVTZ level of theory with density fitting (using default
parameters unless otherwise specified).  Small perturbations (uniform
random noise $\leq$ 0.01 \AA) was added to prevent redundancy in the
dataset.  All coefficients of the relaxed wavefunction were saved in
the Molden format for further analysis in
Multiwavefn,\cite{Multiwfn2024} where ADCH, Becke, CHELPG, CM5,
Hirshfeld, MBIS, MK, and VDD atom centered charges were
obtained. Further, the MBIS multipole expansion ($\ell \leq 4$) was
also derived.  Training used the standard training/validation/testing
splits (8:1:1).\\

\noindent
For models covering \textit{chemical space}, the QM9 database was
employed.\cite{QM9_2014} Data were split into three non-overlapping
training/validation sets based on the order of molecules in the GDB
dataset\cite{GDB_2012}.  A total of 126900 molecules from
QM9\cite{QM9_2014} were split into three folds of 40000 structures
each, and an additional holdout set of 6900 remaining structures was
kept for testing.  These structures were split based on their order in
QM9 (i.e. based on the number of atoms): the training and test set
contained molecules with 1 to 8 and 8 to 9 heavy atoms, respectively.
QM9 consists of molecules containing carbon, hydrogen, oxygen and
fluorine, up to nine heavy atoms.  This introduced a sampling bias due
to the upstream graph enumeration algorithm, leading to chemically
similar structures within individual splits.  Training set sizes of
8k, 16k, and 32k molecules were used to benchmark and, in particular,
learning curves, each with a constant validation set size of 8k
structures. These models were trained for 500 epochs with a batch size
of 16 and a learning rate of $1 \cdot 10^{-4}$. Final models were
trained on 64k structures for 10000 epochs or until the validation loss of the models converged, using an 8k validation set
and 6.9k molecules held out for testing. Molecular similarity was
evaluated using the Tanimoto coefficient computed in
RDKit\cite{rdkit}, based on two-hop Morgan fingerprints (1024
bits). TMAP projections\cite{probst:2020} were generated using the
MHFP encoder\cite{Probst:2018} to visualize molecular space
coverage.\\

\noindent
Analysis of the dataset based on the SMILES string deposited in the
original QM9 dataset was performed using RDKit\cite{rdkit} (version
2023.5) and TMAP\cite{probst:2020}, see Figure \ref{sifig:tmap}.  As is typical when working with
QM9,\cite{Unke:2018,Kim:2019,Weinreich:2021,Unke:2021} structures that
failed simple consistency checks were removed, namely molecules with
energies that were poorly converged or molecules that decomposed
during optimization.  For a given molecule, the energy was converged
at the PBE0/def2-TZVP level of theory using PSI4\cite{PSI4}, and the
wavefunction was analyzed, from which MBIS
multipoles\cite{Verstraelen:2016} up to $\ell=2$ and the electrostatic
potential (evaluated on up to 3200 grid points) were obtained.  Grid
points were sampled on a regular grid generated by increasing the
bounding box of the molecule by $\pm~3.0$ \AA~ along every direction, sampling 15 points along each axis.
The effect of this relatively coarse grid was explored later.
As the electrostatic potential
distorts close to atomic nuclei due to charge penetration effects, a
careful selection of grid points is important.  Different grid point
sampling methods were explored: an element-independent minimum
distance cutoff of 1.7 \AA~ (i.e., $1 \times r_{\rm vdw}$, the vdW radius of the
largest atom, i.e., carbon), and element-dependent minimum distance
cutoffs of $1.4 \times r_{\rm vdw}$. Due to this sampling scheme,
larger molecules typically featured a lower density of grid points.\\

\noindent
Alternative examples of training data with mixed radial and angular
grids were also explored. Specifically, the same grids as used in the
RESP method were tested, \textit{post-hoc} and not used in any data in
the main manuscript; however, the performance was seemingly improved,
based on the quality of the dipole moment prediction.  These
observations are analogous to the long-range Coulomb interaction
problem, which (famously) converges differently depending on how the
long-range interaction pairs are selected.\cite{Kolafa:1992} \\

\noindent
To assess robustness beyond the training setup, the ESP was evaluated
on a finer grid and at a different level of theory. To evaluate the
model's ability to extrapolate, 
reference ESP data was
computed at a different level of theory (MP2/6-31G(d,p)) using
Gaussian16\cite{g16}, in contrast to the DFT-level ESPs used for
training.
Further, ESP predictions were compared to a roughly
ten-fold finer reference ESP grid than was used during training. The
grid was partitioned into three regimes based on the ratio of the separation distance versus van der Waals radii ($r/r_{\rm vdw}$): close-
$(1.20 < r/r_{\rm vdw} < 1.66)$ , mid- $(1.66 < r/r_{\rm vdw} < 2.20)$ , and far-range $(2.20
< r/r_{\rm vdw})$ regions relative to the molecular surface (see Table
\ref{tab:range_data}). 
This ensures a fair assessment of the transferability of the
approach across different levels of theory and grid densities, which is desirable when
employing these models in force-field fitting to alternative reference
data.\\

\noindent
For pretraining on chemical space, additional transfer learning (TL)
experiments were performed outside the scope of the QM9 dataset which
is restricted to equilibrium structures.  Here, coordinates of 2040
random, neutral dipeptides sampled from gas-phase MD simulations at
500 K using were obtained from the SPICE dataset described by Eastman
and coworkers\cite{Eastman:2023}.  Calculations as described above
were performed to obtain the ESPs and molecular dipole moments.

\subsection{Model Training}
The DCM-net models were configured with the following hyperparameters:
two message passing iterations, a cutoff radius of 4.0 Å (used to
define graph edges), and a feature size of 16, and 8 radial basis
functions.  Each feature was expanded up to $\ell = 2$.  These
hyperparameters were kept relatively small for speed of training and
to aid against overfitting; however, increasing the feature size and
number of radial basis functions appears to improve performance, and
can likely be increased without significant drawbacks.  Optimization
of the NN was performed using the ADAM optimizer\cite{ADAM} ($\beta_1
= 0.9$, $\beta_2 = 0.999$, $\epsilon = 10^{-8}$) as implemented in the
Optax library\cite{jax}, with a batch size of one.  The loss function
(Eq. \ref{eq:loss1}) was weighted to balance the ESP and charge
predictions. Specifically, the weight on the electrostatic potential
term, $w_{\rm ESP} = 10^5$ Hartree/$e$.  This choice resulted in ESP
and dipole contributions of comparable magnitude during training, when
the weight for the atomic charges was $w_q = 1.0~e$. Initial experiments used an exponentially decaying learning
rate (from $1 \cdot 10^{-3}$ to $1 \cdot 10^{-5}$) and gradient
clipping (global norm, clip norm = 1)\cite{Pascanu:2013}, with the
objective of minimizing the squared error in the MBIS-derived
monopoles and ESP (see Eq. \ref{eq:loss1}). Learning curves from these
models were used to validate the implementation (Figure \ref{sifig:lcs}, more details are
provided in the SI).  This weighting term was typically 1-2 orders of
magnitude larger for the ESP than the total charge and molecular
dipole terms.  Charge weights $w_q = 1.0~e$ ensured a correct first
moment of the ESP to within 0.001 electron charge units, although
training on datasets containing molecules with a non-zero net charge
may require larger weights.\\

\noindent
When pretraining on chemical space (QM9) and transferring to the
dipeptide dataset, all Ala$_2$ and Gly$_2$ structures (63 conformers)
were deposited in the test set to prevent any data leakage and to
better justify the transferability of this approach. The remaining
structures were used in the training and validation sets (1900 and 77,
respectively).  As in previous TL-studies\cite{Kaser:2021,Kaser:2022},
the learning rate was decreased by a factor of 100 to 2.5e-07, and all
parameters of the model were optimized over 1000 steps of transfer
learning.\\

\section{Results and Discussion}
DCM-net was evaluated along three axes: (1) practical generation of
minimal charge models, (2) ESP accuracy versus MBIS multipoles, and
(3) TL to out-of-distribution molecules. First,
parametrization of conformational space is presented, and the motion
the individual DCM-charges as a function of time and the symmetry of
the models is investigated. Following this, an analysis of the
accuracy of the model is provided in comparison to density derived
atomistic multipole expansions obtained from MBIS, as well as a
discussion on the training dynamics and model validation using the
learning curves.  As a proof-of-concept, minimal distributed charges
are generated for a single molecules (Fluoro-Benzene, FBz) which can be readily used in MD simulations in packages
such as CHARMM.\cite{Hwang:2024} Finally, the capability for applying
these models in a TL-setting is investigated.\\

\subsection{Conformational Space: CO$_2$ and Rotational Equivariance}
To probe how the model represents conformational variability in a
symmetry-sensitive system, carbon dioxide (CO$_2$) is considered. In
order to run dynamics, a message passing neural network as described
previously\cite{Unke:2019} was used to predict energies, forces and
dipole moments via atom centered fluctuating charges. Since the loss
function, Eq. \ref{eq:loss1}, includes a weighted atomic charge
penalty, it is possible to couple the two models during training by
using the learned PhysNet atomic charges as the reference atomic
monopoles $M^{0}_{\rm ref}$ in DCM-net.\\

\begin{figure}
    \centering
    \includegraphics[width=0.95\linewidth]{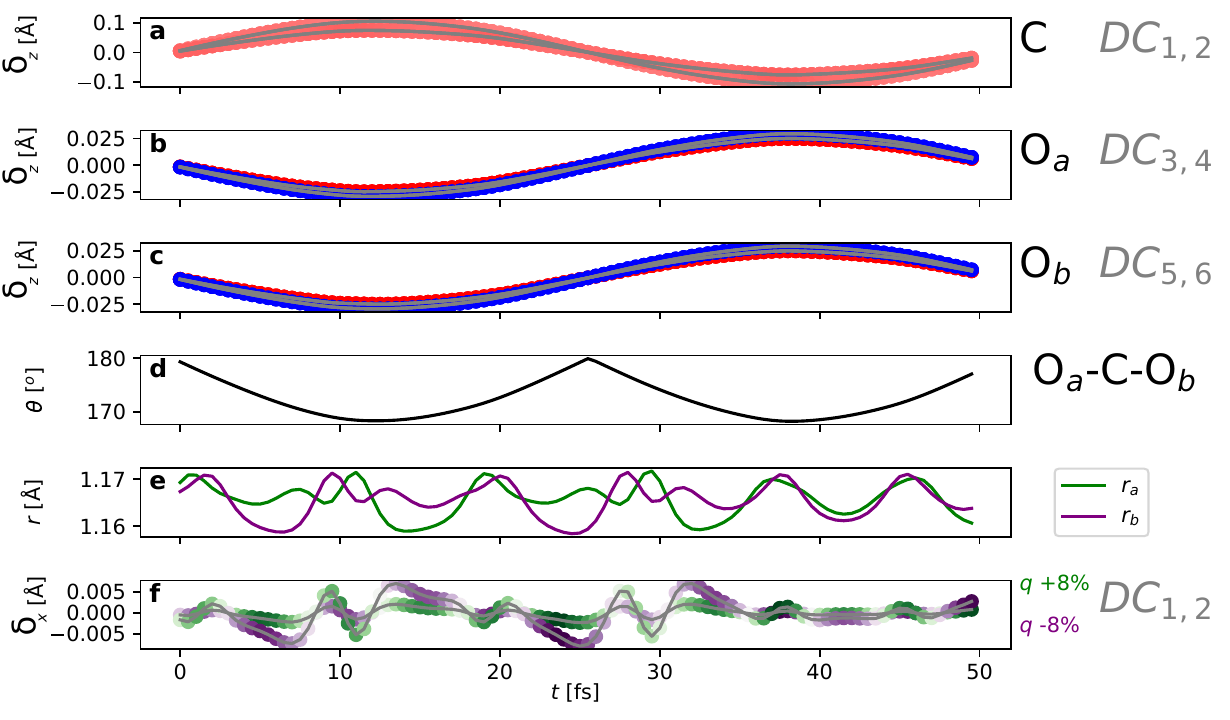}
    \caption{\textit{Conformational space:} Dynamics of the
      DCM-net/PhysNet two charge model of CO$_2$ (a-c) Components of displacements
      for six distributed charges $\mathbf{\delta}$, labeled on the right hand side $DC_{1-6}$, coupled with (d) the
      internal angle $\theta$, shown along the $z$ axis which is
      aligned with bonds (e) $r_a$ and $r_b$. (f) Small displacements
      along the $x$ axis for the central carbon atom distributed
      charges ($DC_{1,2}$), and their magnitude $q$, couples the
      electrostatic potential to changes in bond length. For the
      symmetric case $(\theta =180^{\circ})$, displacements approach
      0.0 \AA, and charges $q$ approach their mean value.}
    \label{fig:co2}
\end{figure}

\noindent
Using the fitted potential energy surface, dynamics simulations were
run. For this demonstration, random initial velocities were assigned
at 300 K and positions were updated using a 0.5 fs time
step. Figure~\ref{fig:co2} shows the dynamics of the DCM-net $n_{\rm
  DC} = 2$ representation as a function of the bond lengths $r_a$,
$r_b$ and the internal angle $\theta$. The distributed charges
associated with each atom respond smoothly to distortions away from
linearity, with displacements occurring primarily along the molecular
$z$ axis, which is aligned with the C--O bonds, see Figure \ref{fig:co2} panels a, b, c, and f. As the molecule
approaches the symmetric configuration $(\theta = 180^\circ)$, these
displacements vanish and the charge magnitudes relax toward their mean
values, consistent with the restoration of inversion symmetry.  In
addition to longitudinal motion, small but systematic displacements of
the carbon-centered distributed charges are observed along the
transverse $x$ direction. These transverse shifts are coupled to
changes in bond length and provide a mechanism for modulating the
electrostatic potential without introducing spurious dipole moments in
the linear limit. This behavior highlights how the model encodes
internal polarization effects through geometry-dependent charge
rearrangements while preserving the correct symmetry constraints. The
resultant models are qualitatively similar to fluctuating distributed
charge models described previously, despite differing in the fitting
approach.\cite{MM.fmdcm:2022, MM.kmdcm:2024} The learned
displacements are smaller for DCM-net, since the magnitude of the
charges is allowed to fluctuate unlike approaches such the
kernel-based minimal distributed charge model.\\

\noindent
In the context of a small molecule potential, it is interesting to
compare the performance of an equivariant (EQV) model versus a
non-equivariant (NEQV) model. A shared parameter budget of 500,000 was
applied to attempt a fair comparison; however, the results are
arguably qualitative.  Table~\ref{tab:metrics-merged} reports test-set
errors for energies, forces, dipole moments, and electrostatic
potentials after 1000 steps of training.  The EQV model yields lower
mean absolute errors for energies, forces, and dipoles compared to the
NEQV model, along with reduced variability across independent training
runs.  Despite these marginal improvements, the performance of the
models was comparable.  Interestingly, the ESP errors were smaller for
the NEQV model than the EQV model (2.59 versus 2.48 (kcal/mol)/$e$) on
the hold-out test set. Seemingly large errors in
Table~\ref{tab:metrics-merged} are related to the highly
non-equilibrium dataset which covers a range over 100 kcal/mol in
energy. Nevertheless, it is worthwhile to note that for such small
  molecules kernel-based methods are often preferable if accuracy in
  the representation of the energies is
  sought.\cite{MM.rkhs:2017,MM.co2:2021}\\

{\small
\begin{longtable}{l *{5}{r}}
\caption{\textit{Conformational space:} CO$_2$. Test Set MAE for energies $E$, forces $\mathbf{F}$, dipole moments $\mathbf{D}$ and electrostatic potential RMSEs $\Delta$ESP,  comparing (non)equivariant, (N)EQV, models. Uncertainties correspond to the standard deviation over three random trials. Large errors in the energy are related to the presence of outlier high-energy structures, as over 100 kcal/mol is covered in the dataset.}\label{tab:metrics-merged}\\
\toprule
 & $E $[kcal/mol] & $\mathbf{F}$ [(kcal/mol)/\AA] & $\mathbf{D}$ [D] & $\Delta\mathrm{ESP}$ [(kcal/mol)/$e$)] \\
\midrule
\endfirsthead
\caption[]{Validation MAE (continued)}\\
\toprule
 & $E$ & $F$ & $D$  \\
\midrule
\endhead
\midrule
\multicolumn{5}{r}{Continued on next page}\\
\midrule
\endfoot
\bottomrule
\endlastfoot

EQV  &
  $1.70 \pm 1.48$ &
  $0.305 \pm 0.19$ &
  $0.122 \pm 0.05$ &
 $2.59 \pm 0.01$ \\

NEQV &
 $1.97 \pm 1.64$ &
 $0.375 \pm 0.25$ &
 $0.127 \pm 0.05$ &
  $2.48 \pm 0.02$  \\

\end{longtable}
}

\noindent
This distinction becomes evident when explicit symmetry consistency
tests, such as random rotations and translations, are
applied. Table~\ref{tab:symmetry-merged} summarizes rotational and
translational errors in dipole moments and electrostatic potentials
under random rigid-body transformations. The EQV model exhibits errors
close to numerical precision, confirming exact rotational and
translational equivariance by construction. In contrast, the NEQV
model shows substantial rotational errors.  Particularly, for the
electrostatic potential, since distributed charges are predicted in 3D
space, without rotational equivariance or averaging through data
augmentation, the variance in errors were roughly 90$\%$ on the test
set.  Such large variations after applied rotations are explained by
the entire dataset containing structures aligned to the principle
molecular axes.\\

\pagebreak

{\footnotesize
\begin{longtable}{l *{4}{r}}
\caption{\textit{Conformational space:} CO$_2$. Rotational and translational variations for dipole moments $\mathbf{D}$ [D] and electrostatic potential RMSEs $\Delta$ESP [(kcal/mol)/$e$)] evaluated on the test set, comparing (non)equivariant, (N)EQV, models. Rotations and translations were sampled uniformly at random.}\label{tab:symmetry-merged}\\
\toprule
 & $\Delta \mathbf{D}_\text{rot}$  & $\Delta\mathrm{ESP}_\text{rot}$  & $\Delta \mathbf{D}_\text{trans}$   & $\Delta\mathrm{ESP}_\text{trans}$   \\
\midrule
\endfirsthead
\caption[]{Symmetry Checks (continued)}\\
\toprule
 & $\Delta\mu_\text{rot}$ & $\Delta\mathrm{ESP}_\text{rot}$ & $\Delta\mu_\text{trans}$ & $\Delta\mathrm{ESP}_\text{trans}$ \\
\midrule
\endhead
\midrule
\multicolumn{5}{r}{Continued on next page}\\
\midrule
\endfoot
\bottomrule
\endlastfoot

EQV  &
  $4.04{\times}10^{-4} $ &
  $0.02$ &
  $6.74{\times}10^{-6} $ &
  $2.26{\times}10^{-4}  $ \\

NEQV &
$5.84{\times}10^{-2}  $ &
 $2.26  $ &
 $3.52{\times}10^{-5} $ &
 $3.06{\times}10^{-3} $ \\

\end{longtable}
}

Across conformational space, these results demonstrate that enforcing
equivariance provides geometrically consistent, and therefore
physically meaningful, electrostatic representations - providing a
basis for more complex models attempting to learn shared
representations across chemical space.

\subsection{Chemical Space: Accuracy of the Electrostatic Potential}
After validating DCM-net for an ensemble of structures of single molecule in the gas phase, an application across the chemical space covered by the QM9 data set is discussed next. For this, DCM-net was trained as described in the Methods section.\\

\begin{figure}[h!]
  \centering
  \includegraphics[width=0.85\textwidth]{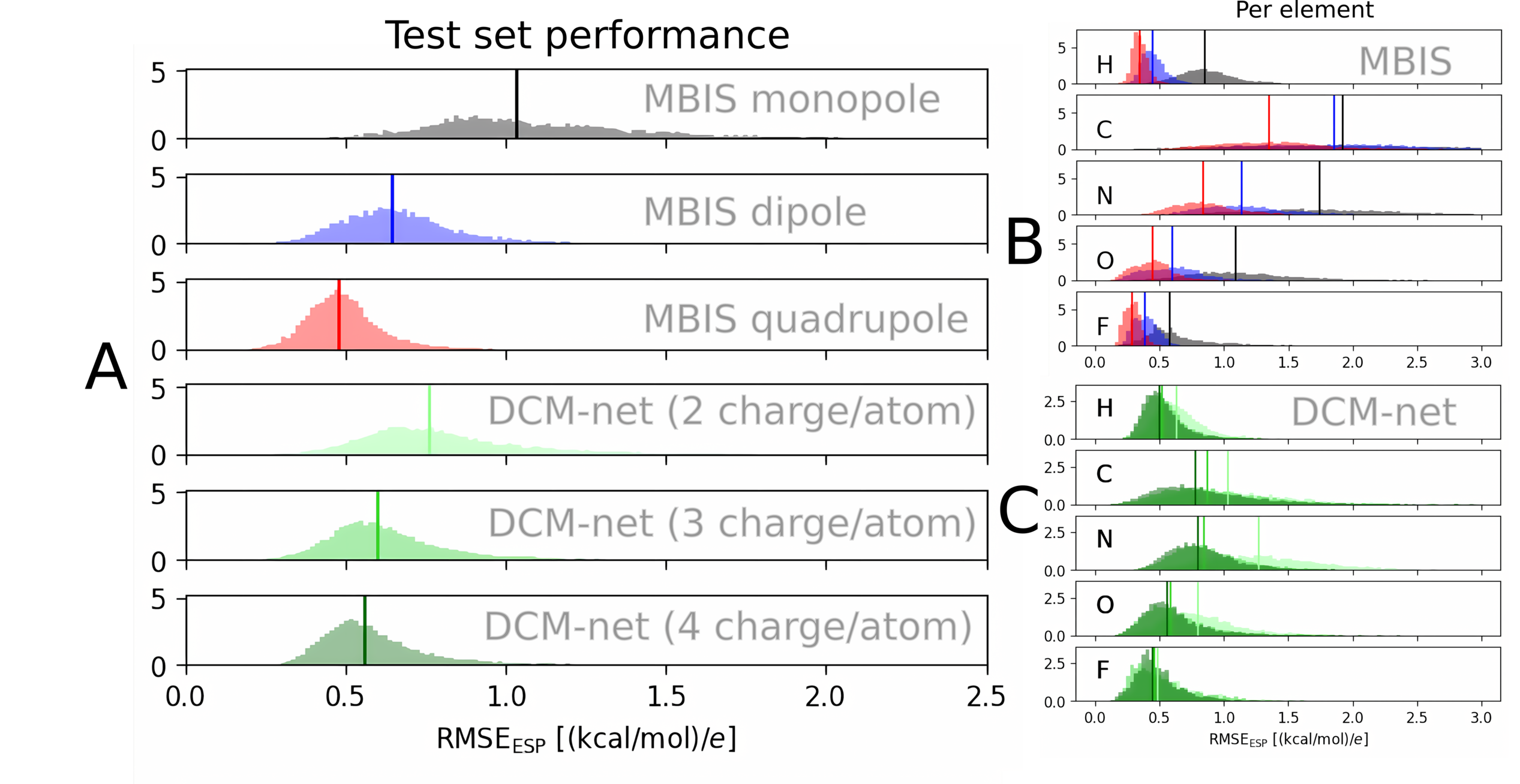}
  \caption{\textit{Chemical space:} QM9 (A) Distribution of
    RMSE$_{\mathrm{ESP}}$ values (difference between the model and
    reference ESP) for the entire hold-out (test) set, (B)
    contributions based on individual elements contributions for
    multipoles and (C) DCM-net.  The normalized probability density is
    shown on the $y$-axis.  Atom centered multipole expansions up to
    monopole (black), dipole (blue) and quadrupole (red) are shown
    along with the two-, three-, and four-charge-per atom models shown
    in pale green, green and dark green, respectively.  The median of
    the distributions are shown with vertical lines. Note the change
    in height of the distributions for the DCM-net models due fatter
    tails in the distribution.}
  \label{fig:esperrors}
\end{figure}

\noindent
The distributed charge positions predicted by DCM-net are able to
reduce the average RMSE error in the ESP in comparison to fitted MBIS
monopoles, and reach similar accuracy of the fitted multipole
expansion up to $l=1$ when more than $n_{\rm DC} > 2$ are
predicted (Table \ref{tab:esp_results}); however, the accuracy for the model with
$n_{\rm DC} > 4$ is slightly worse than the MBIS expansion
up to atomic quadrupoles (0.6 versus 0.5 (kcal/mol)/$e$ on the test
set).  Figure \ref{fig:esperrors}A reports the distribution of ESP
errors for increasing multipole ranks/number of distributed charges
per atom for all molecules in the test set.  The shape of the error
distributions for multipole- and distributed charge-based ESPs are
comparable, indicating a uniform improvement for the test set as the
number of distributed charges increases.  If the error distributions
associated with the predicted distributed charges were skewed with
heavy tails it may suggest that the learned distributed charge
representations were biased towards certain types of molecules;
however, the comparable improvements in the error distributions suggest
the DCM-net approach is transferable (at least within the chemical
space of QM9) and increasing the number of distributed charges
provides benefits almost on par with increasing the rank of the
multipole expansion.\\

\noindent
The sum of the distributed charges per atom closely matches the target
MBIS monopoles (Table \ref{tab:esp_results}) implying that the
improvement is a result of the anisotropic nature of the off-center
charges and not due to the model `cheating' by assigning physically
unrealistic charges to buried atoms (i.e. no accessible volume or grid
points evaluated by the loss function for those atoms). This
assessment is further supported by comparing the distributions of such
charges obtained using MBIS and DCM-net, as well as a general fixed
charge force field such as CGenFF (see SI, Figure \ref{sifig:charges}).
By applying this constraint, predictions from DCM-net can be converted
to an equivalent atom-centered model on an atom-by-atom basis to
create minimal models with acceptable accuracy and low computational
complexity.\\

\noindent
To analyze for which chemically relevant
situations and features using distributed charges is most beneficial,
each ESP-grid point was associated with the atom closest to it in space: either (1)
by element (Figure \ref{fig:esperrors}B and C), or (2) by CGenFF atom-type (Tables \ref{sitab:cgenffMBIS} and \ref{sitab:cgenffDCMNet}).
Although the ESP at a certain grid point contains contributions from
the entire molecule, the closest atom/distributed charge is expected
to have the largest impact on the associated error. Additionally,
since atoms such as carbon are usually buried and occluded by hydrogen
atoms, the relative errors for different elements should not be be
compared directly, and a comparison between different electrostatic
models is more consistent.\cite{Wang:2012, stone:2013} For
atom-centered multipoles, the average quality of the ESP around carbon
and nitrogen atoms is only significantly improved with the addition of
atom centered quadrupoles as many heteroaromatic rings present in the
test set are expected to have significant quadrupole contributions to
the ESP related to the presence of $\pi$-bonds.\cite{stone:2013,
  Day:2005} In contrast, distributed charges appear to be an efficient
way of representing such anisotropic charge distributions as even
the $n_{\rm DC} = 2$ model reaches quadrupolar accuracy around these atomic centers.\\

\noindent
For local environments around fluorine atoms, the improvement gained
for $n_{\rm DC} > $2 is much smaller in comparison to
increasing the rank of the multipole expansion. In contrast, the
addition of atomic quadrupoles nearly halves the median
RMSE$_{\mathrm{ESP}}$.  For oxygen atoms, adding atom centered dipoles
reduces the median RMSE$_{\mathrm{ESP}}$ from $\sim 1.1$ to $\sim 0.6$
(kcal/mol)/$e$. This is related to the improved description of hydroxy
(-OH) groups which have a significant dipolar character. In contrast,
improvement gained from a model with $n_{\rm DC} = 2$ instead is
slightly smaller (from $\sim 1.1$ to $\sim 0.8$ (kcal/mol)/$e$). The
performance of the atom-centered dipole model is reached with $n_{\rm
  DC} = 3$, whereas $n_{\rm DC} = 4$ does not provide further
improvements.\\

\noindent

\begin{figure}[!ht]
\centering
\includegraphics[width=0.85\linewidth]{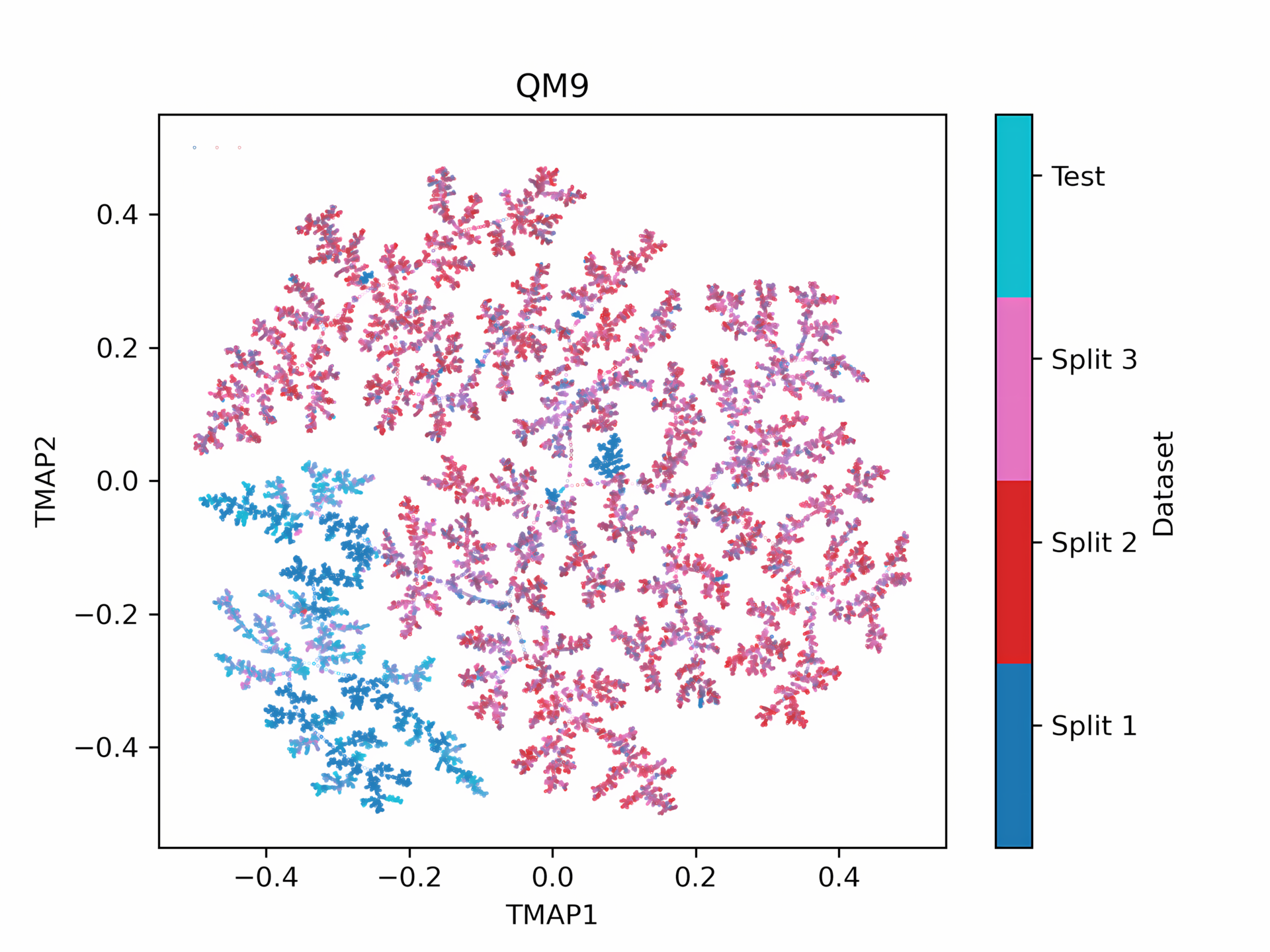}
\caption{\textit{Chemical space: QM9\cite{QM9_2014}. 
TMAP\cite{probst:2020}} projection of the  dataset colored by split. Learning
curves were generated from training on Splits 1 to 3 separately.
Final models used all samples from Splits 1 - 3 in the
training/validation splits. The test set contains atoms with either 8
or 9 heavy atoms. As observed in previous studies\cite{Vazquez-Salazar:2022}, 
splitting by the order of the dataset introduces a sampling bias.}
\label{sifig:tmap}
\end{figure}

\noindent
The same analysis was repeated by separating the grid points into a
larger number of classes based on their proximity to different CGenFF
atom types (see SI, Tables \ref{sitab:cgenffMBIS} and \ref{sitab:cgenffDCMNet}).
Here, only molecules in the test set with available CGenFF atom types were analyzed and the amount of data
for each class is significantly lower and considering the limitations
mentioned above it is more difficult to make well-grounded conclusions
as to which atom types benefit the most from improved electrostatic
representations.  As expected, increasing the rank of the MBIS
multipole expansion monotonically decreases the errors associated with
all atom types (Table \ref{sitab:cgenffMBIS}).  In contrast, using this
ad-hoc partitioning scheme reveals that the learned distributed charge
representations only provide a small improvement in comparison to MBIS
monopoles for certain atom types and, furthermore, the improvement
gain from increasing the number of distributed charges per atom is not
monotonic in all cases likely due to the stochastic nature of training
and the non-unique nature of the fitting problem resulting in a glassy
loss landscape.  Unsurprisingly, given the chemical diversity of the
four splits in the SI (Figure \ref{sifig:tmap}), chemical biases were
observed in training across splits.  \\

\begin{figure}
  \centering
  \includegraphics[width=0.89\linewidth]{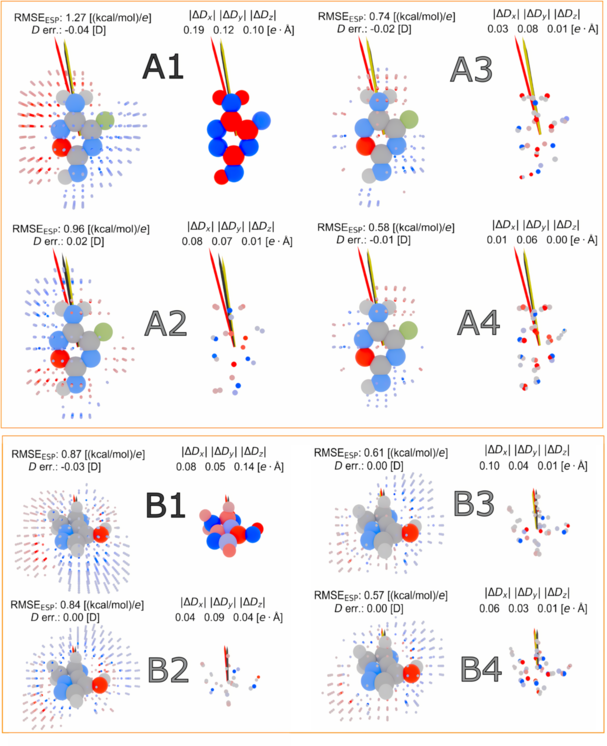}
  \caption{\textit{Chemical space:} The spatial extent of the
    improvement introduced by DCM-net in comparison to MBIS point
    charges (note the color scale is qualitative; identifying regions
    of positive/negative errors and charge), for randomly selected  test
    set examples labeled. Panel A (C$_3$H$_3$N$_4$OF) and panel B (C$_5$H$_7$N$_3$O),
    with (1, black) MBIS point charges, (2-4, gray) DCM-net
    predictions for $n_{\rm DC} = 2, 3, 4$, respectively. On the left, grid points with greater than 75\% of
    the absolute error for the monopole ESP. On the right, molecular
    dipole vectors (scaled by $4\times$ for clarity) of (black) the
    ground truth, and of (red) the MBIS monopole and (yellow) DCM-net
    models. Errors, typically, decrease as the number of distributed
    charges per atom increases. Atoms are colored gray (carbon), white
    (hydrogen), red (oxygen), blue (nitrogen), and green (fluorine).}
  \label{fig:esp-example1}
\end{figure}

\noindent
Figure \ref{fig:esp-example1} shows examples of the positions of the
predicted distributed charges, and grid points of the error surface
with errors higher than 75\% of the mean absolute error obtained with
fitted atom-centered point charges. The error is
expected to be larger at points closest to the molecule where the potential is large; grid
points further away with significant errors ($\geq$75\% of the baseline error) point to regions where anisotropy can be improved.
Strikingly, all distributed charge models reduce the error on
grid points around the fluorine atom (shown in green, Figure
\ref{fig:esp-example1}A2-A4), which is expected to have pronounced
anisotropic features in the ESP due to the $\sigma$-hole that is known
to be difficult to capture with atom centered point charges
alone.\cite{Ibrahim:2011,schyman:2012,MM.mtp:2016,Yan:2017} The
location of grid points have implications seen in the predicted charge
magnitudes.  For instance, for models trained on grid points closer
than $1.4 \times r_{\rm vdw}$ of the atoms (i.e. points where
the ESP is influenced by charge penetration effects), the
spatial distributions of grid points with high errors were typically
larger, although the DCM-net models had lower overall RMSEs than the
MBIS-fitted point charges. This demonstrates a bias towards
close-range points of the ESP. After addressing the issue of the selection of the grid points on which to evaluate the ESP in the training data, e.g. by excluding close range point, the spatial
distributions of errors were comparable to the MBIS-fitted point
charges without any drawbacks to the overall accuracy of the ESP.\\

\noindent
As for all deep-learning models, performance drops are expected when
leaving the training distribution.\cite{Vazquez-Salazar:2021,
  Vazquez-Salazar:2022, yang:2024} One limitation of the machine
learned distributed charges is the presence outliers in the of
long-tails of the distribution.  For some structures, the error in the
ESP is significantly larger than that of the equivalent MBIS
monopoles.  An important consideration in the extrapolation capability
of atomistic neural network models is the coverage of chemical in the
training data space\cite{Vazquez-Salazar:2021, Vazquez-Salazar:2022}
(see SI, Figure \ref{sifig:tmap}).  The maximum Tanimoto
similarity\cite{tanimoto1958elementary} of the 100 structures with
highest error in the test was 0.4; compared to 0.5 for 100 structures
with the lowest errors, which suggests that these outliers may be
related to insufficient training examples for particular classes of
molecules.\\

\begin{table}[t!]
  \centering
  \begin{tabular}{llrr}
    \toprule
    \toprule
    &   $\langle$RMSE$_{\mathrm{ESP}}$$\rangle$ & MAE$_{l=0}$ & MAE$_{D}$\\
    \hline
    \bottomrule
    \multirow[t]{2}{*}{$l=0$} &  1.1 & - & 0.195 \\
    \bottomrule
    \multirow[t]{2}{*}{$l\leq1$} &  0.7 & - & 0.001 \\
    \hline
    \multirow[t]{2}{*}{$l\leq2$} &  0.5 & - & 0.001 \\
    \cline{1-4}
    \cline{1-4}

    \multirow[t]{2}{*}{$n_{\mathrm{DC}}$=2}  & 0.8 & 0.004 & 0.123 \\
    \cline{1-4}
    \multirow[t]{2}{*}{$n_{\mathrm{DC}}$=3}  & 0.7 & 0.003 & 0.110 \\
    \cline{1-4}
    \multirow[t]{2}{*}{$n_{\mathrm{DC}}$=4}  & 0.6 & 0.003 & 0.105 \\
    \cline{1-4}
    \bottomrule
  \end{tabular}
  \caption{\textit{Chemical space:} Performance of the DCM-net models
    with $n_{\mathrm{DC}}$ distributed charges per atom, measured by
    RMSE$_{\mathrm{ESP}}$ error with respect to the reference ESP, in
    comparison to MBIS multipoles. The average RMSE of the ESP for the
    molecules in the test set $\langle$RMSE$_{\mathrm{ESP}}$$\rangle$
    values are in (kcal/mol)/$e$.  The mean absolute error between the
    sum of the DCM-net distributed charges per atom and MBIS
    MAE$_{l=0}$, MAE$_{l=0}$, is in elementary charge $e$, and the
    mean absolute error of the components of the molecular dipole
    moment, MAE$_{D}$, is given in Debye.}
  \label{tab:esp_results}
\end{table}

\noindent
The mean absolute errors of the components of the molecular dipole
moment for various models are given in Table \ref{tab:esp_results}.
Although the $n_{\rm DC} = 2$ model significantly reduces the error in
the molecular dipole moment in comparison to MBIS monopoles (from
0.195 D to 0.123 D), the improvement saturates with increasing numbers
of distributed charges per atom with the three and four charge models
exhibiting only slight improvements (0.110 D and 0.105 D). These
results are inline with previous atom centered point charge models
such as PhysNet\cite{Unke:2019}; however, still behind that of atom
centered multipoles\cite{Thrlemann:2022} and naturally behind the
performance of MBIS atomistic dipoles derived from wavefunction based
densities so naturally fit the data better.\cite{Verstraelen:2016} In
contrast, the MBIS atomic dipoles are, by construction, functions
of local atomic densities and are fit to reproduce the total dipole of
the molecular wavefunction and as a result the associated errors are
small (0.001 D).  A significant advantage of using atomic dipoles
is that unlike dipoles obtained from two off-center point charges, atom centered dipoles can simply be added together to obtain
the molecular dipole moment since the two are both rank-1 tensors.\\

\subsection{Generating Minimal Distributed Charge Models}
\begin{figure}
  \centering
  \includegraphics[width=0.75\linewidth]{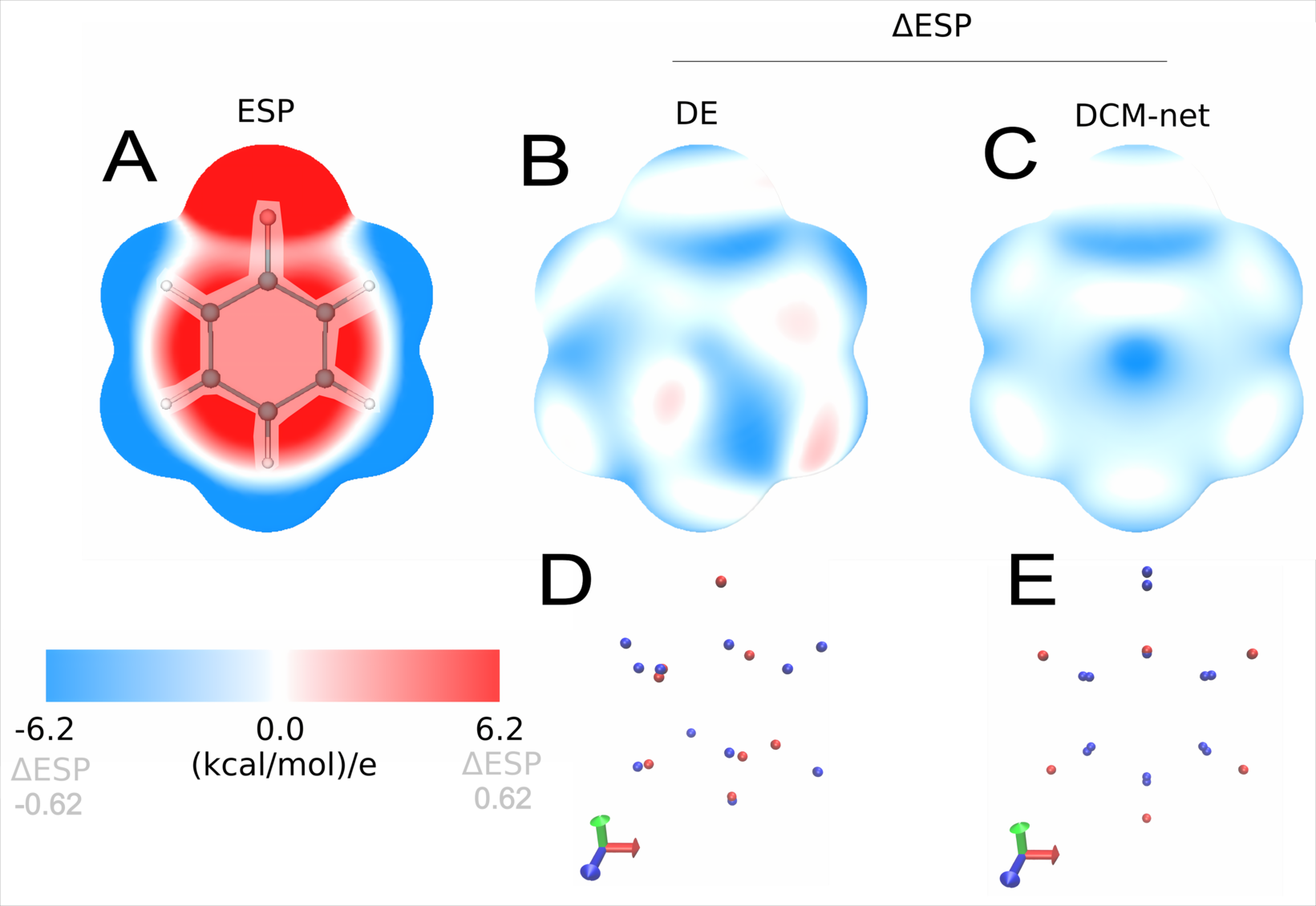}
  \caption{\textit{Model building:} Fluorobenzene (F-Bz). The ground truth
    electrostatic potential surface of using a color map from -6.2 to
    (6.2 kcal/mol)/$e$, at an isodensity of 0.01 a.u., compared to the
    error surfaces from a 19 charge model generated by differential
    evolution (DE) and DCM-net ($n_{\rm DC} = 2$), using a color map
    from -0.62 to 0.62 (10\% of the original color scale). Both models
    achieve comparable accuracy (0.6 versus 0.5 (kcal/mol)/$e$). F-Bz was not present in the training set and serves as a rigorous prediction test for DCM-net.
    The error surface
    of the equivariant model is consistent with the symmetry of the
    molecule.}
  \label{fig:fbenz}
\end{figure}

The model's ability to generate ``minimal'' distributed charge models
was investigated.  Fluorobenzene (Figure \ref{fig:fbenz}) was selected
for this task, as the molecule was not included in the training set
but is, importantly, still within the data distribution of QM9 (i.e.
$<10$ heavy atoms).  The results are compared with a model obtained
using a differential-evolution optimization scheme\cite{MM.dcm:2017}
where 19 charges were selected as a viable trade-off between
complexity and accuracy.  Having established that DCM-net produces
symmetry-consistent minimal models comparable to DE, the ESP accuracy
against baselines from atom-centered multipoles is assessed next. In
this case, the model was obtained from $n_{\rm DC} > 2$ per heavy atom, whereas hydrogen atoms were assigned one on-site
charge. The model obtained from
DCM-net was similar or slightly better than the model obtained from DE
for all regions of the ESP, including maximum errors (Table
\ref{tab:range_data}).\\

\noindent
The DCM-net model leverages the symmetry of the local molecular
environment (due to message passing and the inclusion of rigorous
equivariance), and as a result, the distributions of errors across the
ESP surface are also symmetric. 
It is
also possible to enforce symmetric charge positions based on the
entire point group of the molecule; however, such high symmetry
examples become less common for larger molecules. Leveraging the local
symmetries of the atomic environments is clearly advantageous, as it
leads to similar distributed charge positions for similar functional
groups, without any requirements on the molecule global symmetry, and
can smoothly interpolate between similar molecules/conformations.
Conformational averaging (i.e. finding charge distributions which are
consistent across different molecular conformations) is done
implicitly through the selection of the training data, allowing for
the model to better generalize to new conformations.  In contrast, the
DE routine requires the optimization of the charge positions for each
conformation.  Given these advantages, a practical application of this
model is in automated force field and topology
builders.\cite{stroet:2018,kong:2024,feng:2023,Hwang:2024,gromacs:2024}
\\

\begin{table}
  \centering
  \begin{tabular}{lcccc|c}
    \hline
    & \multicolumn{2}{c}{\textbf{DCM-net}} & \multicolumn{3}{c}{\textbf{DE}} \\
    \textbf{Range} & RMSE$_{\mathrm{ESP}}$ & max. &
    RMSE$_{\mathrm{ESP}}$ & max. & $N_{\mathrm{grid}}$ \\
    \hline
    Total & 5.4e-01 & 12.8  & 6.2e-01 & 12.9 & 415153 \\
    \hline
    Close $(1.20 < r < 1.66)$ & 1.1 & 12.8  & 1.2 & 12.9 & 90064 \\
    Mid-range $(1.66 < r < 2.20)$ & 2.3e-01 & 1.4  & 3.3e-01 & 1.5 & 152428 \\
    Far-range $(2.20 < r)$ & 7.5e-02 & 0.4  & 1.1e-01 & 0.6 & 172661 \\
    \hline
  \end{tabular}
  \caption{\textit{Model building:} Fluorobenzene. Comparison of RMSE, Max Error, and Number of Points for
    two sets of data across different ranges. The grid points are
    selected based on $r$, the ratio of the distance divided by the vdW
  radii for the closest atom.}
  \label{tab:range_data}
\end{table}

\subsection{Transfer Learning to Non-Equilibrium Structures}
Finally, generalization to flexible, protein-like motifs is considered
by adapting the model trained in 3.2 on QM9 to a set of dipeptides via lightweight transfer
learning. A strict training set containing unseen dipeptides (Ala$_2$
and Gly$_2$) was enforced to see if the information contained in QM9
and the remaining neutral dipeptides, would allow the model to
generalize out of distribution.  Transfer learning exploits mutual
information between different tasks to speed-up training and to
increase performance in downstream (usually orthogonal)
tasks.\cite{Kaser:2021,Kaser:2022} Here, TL-predictions of the model
based on QM-minimized structures to structures obtained from cruder
molecular mechanics-based methods is investigated.\\

\noindent
Training on
structures from QM9 introduces a bias to (near) equilibrium
conformations\cite{QM9_2014,Kim:2019} One meaningful test of the consistency of the model with respect to non-equilibrium conformations carried out was to take the SMILES representation of
the molecule and recalculate the molecular geometry using RDKit's
standard (empirical) distance-geometry-based conformer
generator\cite{rdkit}. These structures were typically tens of
kcal/mol higher in energy than the geometries obtained from QM9, and
contain molecules with substantially different distributions of bond
lengths and angles present in QM9.\\

\noindent
In this out-of-distribution regime, the ESP errors were generally
larger than those of the fitted MBIS monopoles. These are not deficits
of the model \textit{per se}, since the data set can be adapted and
the model can be retrained using a
TL-strategy\cite{Kaser:2021,Kaser:2022}. The final average ESP error
over the unseen structures was 0.6 (kcal/mol)/$e$, which is a 0.2
(kcal/mol)/$e$ improvement over the MBIS monopoles (0.8
(kcal/mol)/$e$); and the MAE of the molecular dipole was 0.1314 D in
comparison to 0.2552 D.\\

\noindent
Modest improvements in comparison to fitted MBIS monopoles are
consistent with the results in Table \ref{tab:esp_results}, and
suggest that the $n_{\rm DC} = 2$ model is not competitive with higher
order multipolar expansions derived from atomic densities (bear in
mind these are fit to reproduce the first three moments of the ESP),
as the inclusion of atomistic dipoles with and without quadrupoles
achieved 0.4 and 0.3 (kcal/mol)/$e$ in the average RMSE of the ESP,
and reached a MAE of 0.002 D across predictions of the molecular
dipole moment.  The weaker results on the dipole moment are due to (1)
approximation errors due to using point charges (instead of atomistic
dipoles) and (2) difficulties in training with multiple objectives and
constraints in the loss function. Training iteratively between putting
higher weights on either the dipole or ESP objectives, improved the
final results slightly, as is observed when training models on energy
and forces.\cite{Unke:2024} During transfer learning, the observed
decrease in required training time and increase in accuracy constitute
additional evidence that the features learned by the model are robust
and transferable to a diverse set of molecular conformations or
chemical systems.

\section{Conclusions}
Across tasks, DCM-net produces symmetry-consistent minimal models
quickly and reduces ESP error relative to fitted MBIS monopoles,
approaching dipole-level multipolar accuracy when two or more charges
per atom are used, particular in regions with $\pi-$bonds and
$\sigma$-holes.  Equivariance through construction, as
presented here, is ultimately a design choice and alternatives such as
data augmentation are routinely employed.\cite{Bronstein:2017,
  Pozdnyakov:2023} The approach outlined in the present work streamlines the
generation of distributed charges with  specified accuracy or
complexity, depending on the chosen $n_{\rm DC}$.  These models can be introduced as rigid distributed point
charges into empirical force fields and given a diverse training set,
additional parameters can be optimized against a consistent charge
model in a transferable fashion, allowing for the rapid development of
force fields with anisotropic electrostatics.  Furthermore, studies of models that include conformational sampling for the training data suggest that distributed charges can be
used alongside machine learned potential energy surfaces to improve
the description of polarization and electrostatics.

\begin{acknowledgement}
  This work was supported by the Swiss National Science Foundation
  through grants $200020\_219779$ and $200021\_215088$ and the
  University of Basel. 
\end{acknowledgement}

\begin{suppinfo}
  All data and computer codes for this work are available from the authors upon reasonable request.
  A demo of the DCM-net model is
  available as a Space on HuggingFace at
  \url{https://huggingface.co/spaces/EricBoi/DCMNet} which sources the publically available repository at \url{https://github.com/EricBoittier/dcmnet}.
  The core functionality for building the training datasets, as well as training and evaluating the combined DCM-net and PhysNet model are available from the public repository at
 \url{https://github.com/EricBoittier/mmml.git}. 
  

\end{suppinfo}

\bibliography{references}

\clearpage

\section{Why Equivariance?}

\begin{figure}
\centering
\includegraphics[width=0.75\linewidth]{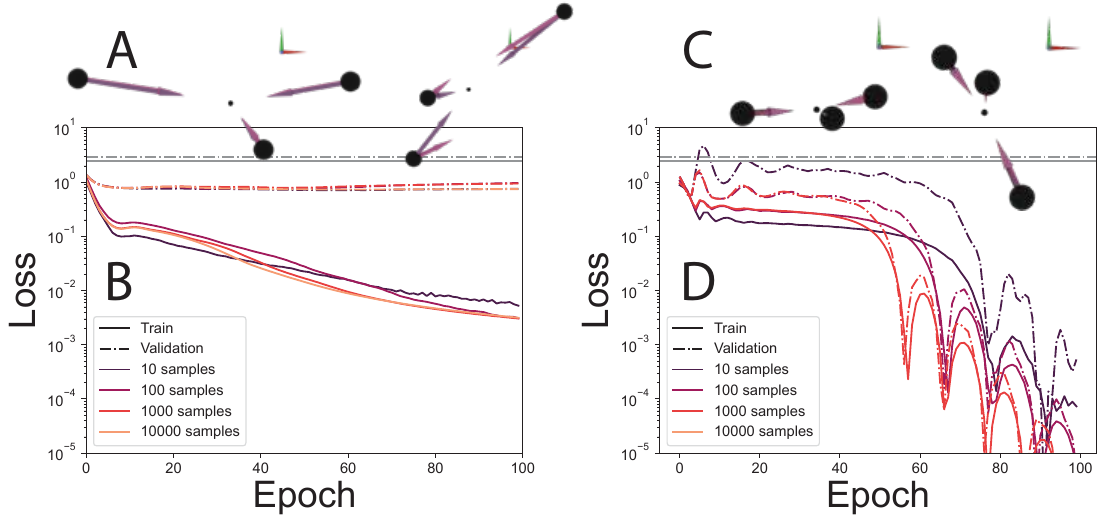}
\caption{(A) Consider a simplified setting where the task is to
  predict atom-centered vectors pointing to the center of mass. (B) A
  feedforward neural network shown examples exclusively in the
  $xy$-plane is unable to generalize if a random rotation is applied to
  the validation set. (C) Prediction errors from equivariant neural
  networks (ENNs) are agnostic to global orientation. (D) As a result,
  ENNs can be more accurate and sample efficient when data is limited
or training is costly.}
\label{sifig:example}
\end{figure}

\begin{figure}
\centering
\includegraphics[width=0.8\linewidth]{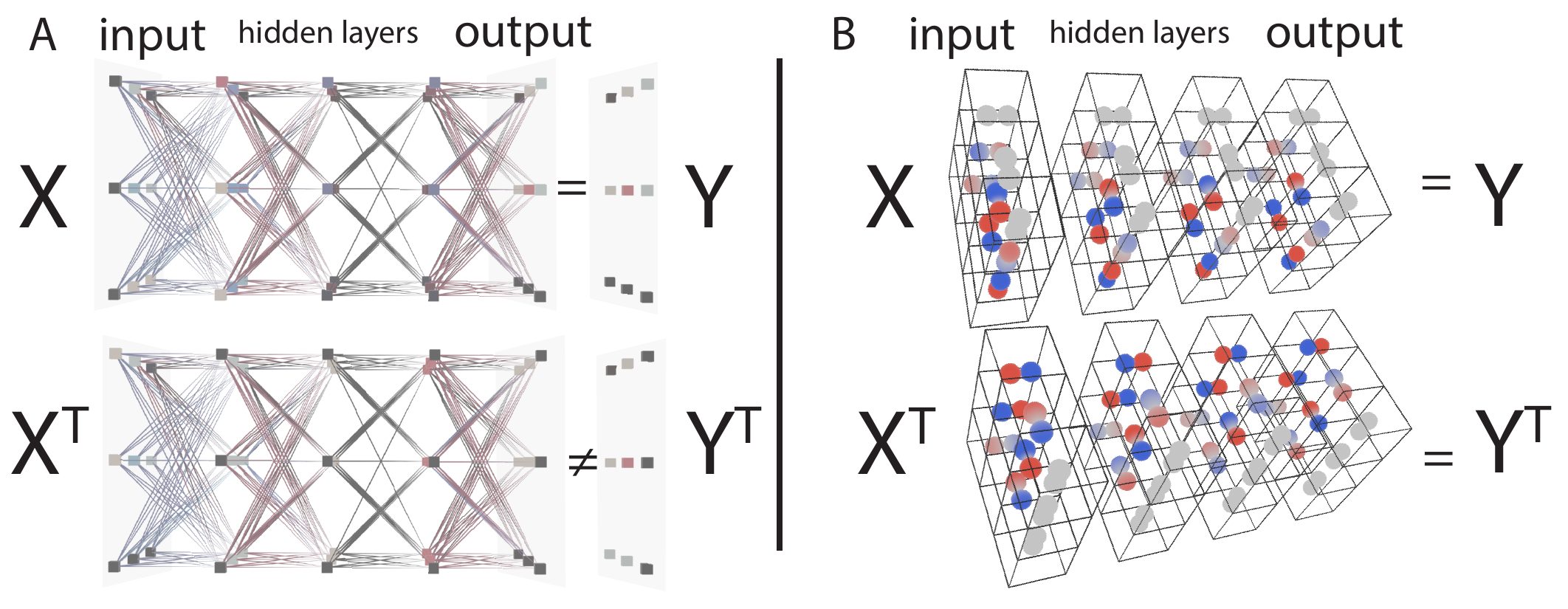}
\caption{The input, hidden state, and output of a three-layer fully
connected network with nodes and edges/weights colored by value for a
NN (A) and (B) an ENN (only the $\ell_p=1_-$ features are depicted
and the weights, i.e. coupling paths, are not shown for clarity). If
the input $X$ and output $Y$ is transposed, the NN prediction is
incorrect because the weights are insensitive to this transformation.
For the ENN, since all operations are equivariant, the hidden states
and output are equivalently transposed when shown $X$ and $X^T$.}
\label{sifig:hiddenstates}
\end{figure}

\noindent
The advantage of equivariance is motivated with a simplified example, working in 3D space.\cite{Bronstein:2017}
Consider a regression task aimed at predicting a set of scaled vectors pointing to the center of mass of three points (Fig. \ref{sifig:example}A).
Here, a non equivariant three-layer feed-forward network with a hidden feature size of 21 and a two-layer equivariant network with a feature size of 3 are trained on points positioned in the $xy$-plane (Fig. \ref{sifig:example}). 
A validation set is constructed where the points can take arbitrary positions in 3D space.
The standard neural network is unable to generalize to the validation set (Fig. \ref{sifig:example}B) because it did not see these rotated examples. 
In an ENN, the inputs in the training and validation sets are equivalent (albeit rotated) so the test and validation accuracy is comparable, and considerably lower than the standard NN in this toy example.\cite{Schtt:2021,Batatia:2022}
Visualizing the hidden states of these networks provides some intuition to this different behavior (Figure \ref{sifig:hiddenstates}). 
If the input $X$ is permuted via a transpose to $X^T$ (which corresponds to a valid rotation) the expected output $Y$ should also be permuted to $Y^T$. 
However, the learned weight matrices of the NN cannot be permuted without causing the output to become incorrect (Figure \ref{sifig:hiddenstates}A).
In the equivariant example, if all $l=1$ features of the input, hidden states, and output are visualized, the states of the neural network appear equivalent for $X$ and $X^T$ and only differ by transposition (Figure \ref{sifig:hiddenstates}B).
If data augmentation is applied by randomly rotating examples, the standard neural network is able to learn this simple function eventually; however, more data and longer training times are usually required to achieve results on par with ENNs.\cite{Pozdnyakov:2023}
Although this example is extremely contrived, it is common for quantum chemistry software to rotate a molecule to a `standard orientation' (aligned to the principal axes of inertia) so such biases illustrated in the training/validation curves in Fig. \ref{sifig:example}A) may be introduced if data augmentation is not applied.\cite{Zare:1988}\\

\noindent
Considering the training loss (solid lines in Figures
\ref{sifig:example}B and D), both NNs perform comparably although the
ENN reaches considerably lower values. However, for the validation
set (dashed lines), the non-ENN is unable to reach comparably low
loss values whereas for the ENN the loss on training and validation
sets is comparable. 
If data augmentation is restrictive, for example due to training times, 
biases illustrated in
the training/validation curves in Fig. \ref{sifig:example}A). 
In other words, using equivariant NNs may be a data reduction strategy.\\

\section{Training Dynamics and Learning Curves}
After initialization and even after the first training epoch, the
predicted displacement vectors of the distributed charges have
approximately zero magnitude, resulting in atom-centered charges.
Atom centered charges are equivariant by definition and are a local
minimum for the model. To elaborate, finding suitable displacements
that generalize across the training set is difficult and atom
centered solutions are a simple way to reduce the loss function early
on during training.
It was observed that, when using gradient clipping, constraining
charge displacements and a small learning rate, it was possible to
trap the models in this `atom-centered local minimum' regime and
subsequently converge the models to an accuracy on par with MBIS monopoles.
However, by restarting training without gradient clipping or
constraints on charge displacements, the models were able explore
true distributed charge solutions and escape such local minima, after
which the training could be restarted again with the original protocol.\\

\begin{figure}[!ht]
  \centering
  \includegraphics[width=0.65\linewidth]{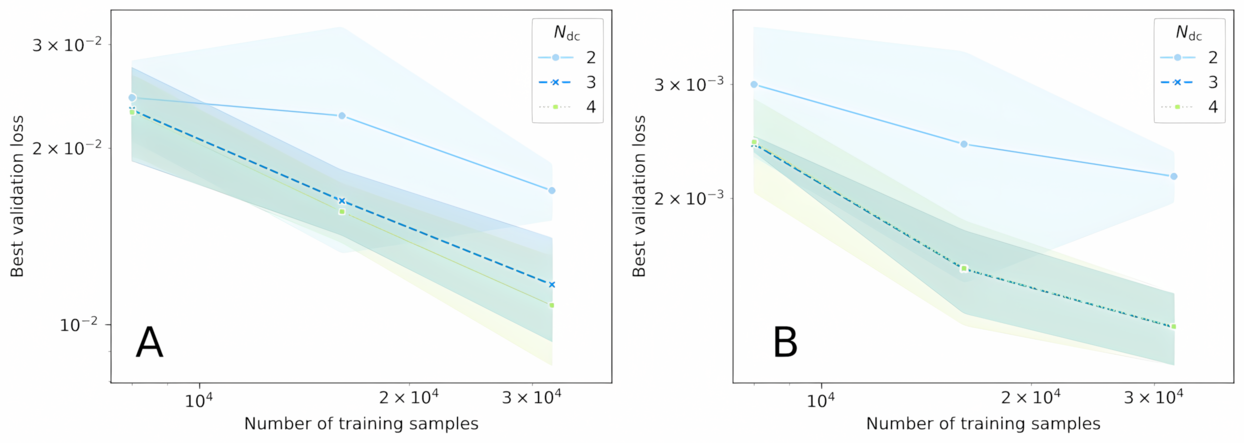}
  \caption{Learning curves for shorter training runs (500 epochs) with
  an ESP weight $w$ of 10$^{4}$ (A) and 10$^{3}$ (B) using three
  non-overlapping splits of the data. Shaded areas report the standard
  deviation for three splits of the training data. Different colors/line 
  styles are for different values of $n_{\rm DC}$: between 2-4 distributed charges per atom.}
  \label{sifig:lcs}
  \end{figure}
  
\noindent 
An important approach in validating new approaches in
deep learning is to verify the improvement of a model given more data, assuming 
there are no redundancies or conflicting information
 in the extended dataset\cite{Vazquez-Salazar:2022}.
According to statistical learning theory, a model should become more
accurate when shown increasing amounts of data.
This process exhibits a power-law
relationship between
the size of the data set and the accuracy of the model, which is also
this case with DCM-net (Figure \ref{sifig:lcs}).
The slope of the learning curves for models trained with
$w_{\mathrm{ESP}}$ of $10^{3}$ and $10^{4}$ Hartree$/e$ are consistent when this
prefactor is accounted for.
This indicates that fitting to the ESP grid is a viable strategy for
obtaining transferable distributed charge models.
The variance in the learning curves is significant (Figure
\ref{sifig:lcs}) due in part to the fact the models were trained on
non-overlapping splits.
The similar scaling performance observed for 3 and 4 charge models
may be related to insufficient training, as the models were not fully
converged after 500 epochs.
The two charge models showed comparable validation loss with that of
the three and four charge models in the low data regime ($< 20$k
training samples) which demonstrates the variance in the performance
of the models.
This observation reinforces the point that finding a set of
distributed charges that reproduces the ESP is non-unique.\\

\clearpage
\section{QM9: Training Data Distributions and Chemistry}
\begin{figure}[h!]
\centering
\includegraphics[width=1.\textwidth]{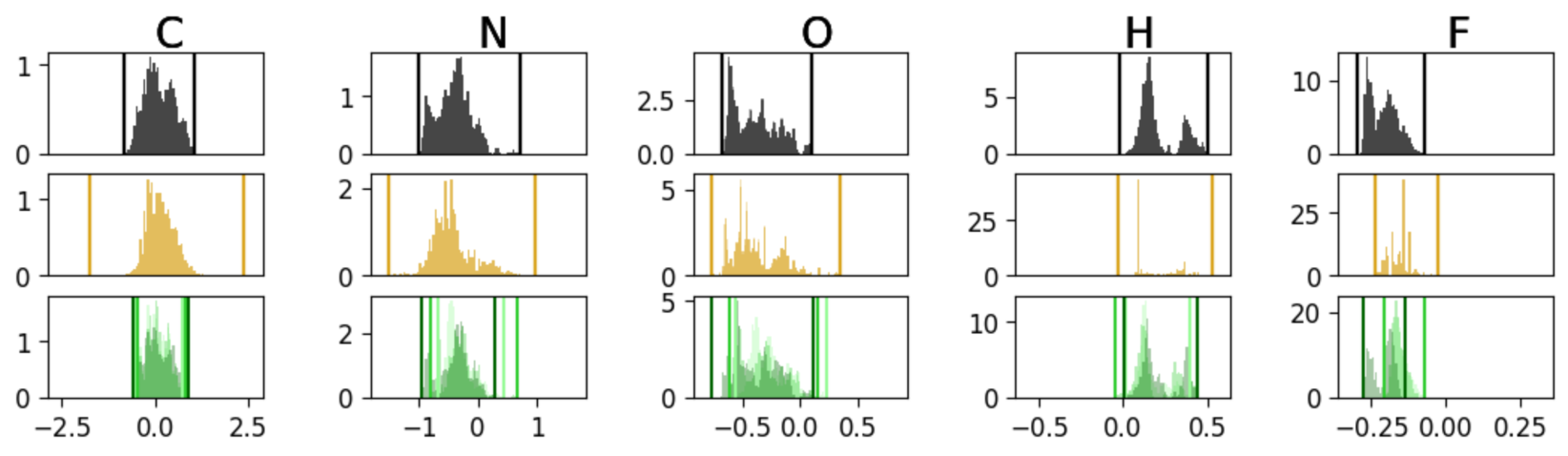}
\caption{Distribution of point charge magnitudes ($e$) for the test set using MBIS
(black), CGenFF (gold), and DCM-net (green). Vertical bars show the range of the charge values.}
\label{sifig:charges}
\end{figure}

Following on from the main text,
more details on the training data distributions and chemistry are provided.
Figure \ref{sifig:charges} shows the distribution of point charges for the test set using MBIS (black), 
CGenFF (gold), and DCM-net (green).
The distribution of charges for the DCM-net model is similar to that of MBIS, indicating that the model 
is able to learn the correct charges for the test set.\\

\noindent
Table \ref{sitab:cgenffMBIS} shows the average RMSE$_\mathrm{ESP}$ ((kcal/mol)/$e$) evaluated on grid points 
grouped by proximity to the nearest CGenFF atom type using fitted MBIS multipole expansions.
Errors less than 0.5 (kcal/mol)/$e$ are shown in bold.
The average RMSE$_\mathrm{ESP}$ for the DCM-net model is similar 
to that of MBIS, indicating that the model is able to
 learn the correct charges for the test set.\\

\begin{table}
\caption{The average RMSE$_\mathrm{ESP}$ ((kcal/mol)/$e$) evaluated
on grid points grouped by proximity to the nearest CGenFF atom type
using fitted MBIS multipole expansions. Errors less than 0.5
(kcal/mol)/$e$ are shown in bold.}
\label{sitab:cgenffMBIS}
\centering
\scalebox{0.7}{\begin{tabular}{lllll}
\toprule
& $l=0$ & $l\leq1$ & $l\leq2$ & description \\
\midrule
CG1N1 & 1.34 & 0.71 & \textbf{0.48} & Carbon in cyano group \\
CG1T1 & 1.42 & 0.61 & \textbf{0.42} & Alkyne carbon (R-C) \\
CG1T2 & 0.84 & \textbf{0.38} & \textbf{0.25} & Alkyne carbon (H-C) \\
CG2R71 & 2.59 & 0.64 & \textbf{0.42} & 7-membered aromatic carbon \\
\hline
FGA3 & \textbf{0.20} & \textbf{0.06} & \textbf{0.04} & Aliphatic
fluorine, trifluoro group \\
FGR1 & \textbf{0.50} & \textbf{0.16} & \textbf{0.10} & Aromatic fluorine \\
\hline
HGA1 & \textbf{0.50} & \textbf{0.12} & \textbf{0.08} & Aliphatic proton, CH \\
HGA2 & \textbf{0.40} & \textbf{0.11} & \textbf{0.07} & Aliphatic proton, CH2 \\
HGA3 & \textbf{0.31} & \textbf{0.09} & \textbf{0.06} & Aliphatic proton, CH3 \\
HGP1 & \textbf{0.49} & \textbf{0.12} & \textbf{0.08} & Polar hydrogen \\
HGP4 & \textbf{0.43} & \textbf{0.09} & \textbf{0.06} & Polar
hydrogen, neutral conjugated -NH2 group \\
HGPAM1 & \textbf{0.49} & \textbf{0.17} & \textbf{0.11} & Polar
hydrogen, neutral dimethylamine group \\
HGPAM2 & \textbf{0.37} & \textbf{0.08} & \textbf{0.05} & Polar
hydrogen, neutral methylamine group \\
HGR51 & \textbf{0.47} & \textbf{0.12} & \textbf{0.08} & Nonpolar
hydrogen, neutral 5-membered planar ring carbon \\
HGR52 & \textbf{0.47} & \textbf{0.13} & \textbf{0.09} & Aldehyde
hydrogen, formamide hydrogen (RCOH) \\
HGR61 & \textbf{0.45} & \textbf{0.16} & \textbf{0.11} & Aromatic hydrogen \\
HGR62 & \textbf{0.50} & \textbf{0.16} & \textbf{0.11} & Nonpolar
hydrogen, neutral 6-membered planar ring \\
\hline
NG1T1 & \textbf{0.31} & \textbf{0.18} & \textbf{0.12} & Nitrogen in
cyano group \\
NG2D1 & 1.09 & \textbf{0.29} & \textbf{0.20} & Nitrogen in neutral imine \\
NG2R50 & 0.66 & \textbf{0.16} & \textbf{0.11} & Double-bonded
nitrogen in neutral 5-membered planar ring  \\
NG2R60 & 0.79 & \textbf{0.27} & \textbf{0.18} & Double-bonded
nitrogen in neutral 6-membered planar ring \\
NG2R62 & 0.77 & \textbf{0.22} & \textbf{0.15} & Double-bonded
nitrogen in 6-membered planar ring with heteroatoms  \\
NG321 & 1.76 & \textbf{0.37} & \textbf{0.26} & Neutral methylamine nitrogen \\
\hline
OG2D1 & 0.56 & \textbf{0.16} & \textbf{0.11} & Carbonyl oxygen \\
OG2D4 & 0.72 & \textbf{0.17} & \textbf{0.12} & 6-membered aromatic
carbonyl oxygen  \\
OG2R50 & 0.76 & \textbf{0.19} & \textbf{0.13} & Oxygen in furan ring \\
OG301 & 1.49 & \textbf{0.37} & \textbf{0.25} & Ether oxygen (-O-) \\
OG311 & 0.84 & \textbf{0.18} & \textbf{0.12} & Hydroxyl oxygen \\
OG3R60 & 1.01 & \textbf{0.31} & \textbf{0.21} & Oxygen in 6-membered
cyclic enol, ether, or ester \\
\bottomrule
\end{tabular}}

\end{table}

\clearpage
\begin{table}[]
\caption{The average RMSE$_\mathrm{ESP}$ ((kcal/mol)/$e$) evaluated
on grid points grouped by proximity to the nearest CGenFF atom type
using trained DCM-net models with the best performance on the training
set. Errors less than 0.5 (kcal/mol)/$e$ are shown in bold.}
\label{sitab:cgenffDCMNet}
\centering
\scalebox{0.7}{\begin{tabular}{lllll}
\toprule
& $n_{dc} = $$2$ & $ 3$ & $ 4$ & description \\
\midrule
CG1N1   & 1.30        & 1.13       & 1.16 & Carbon in cyano group  \\
CG1T1   & 1.45        & 1.12       & 1.16 & Alkyne carbon (R-C) \\
CG1T2   & 0.80        & 0.70       & 0.63 & Alkyne carbon (H-C) \\
CG2R71  & 2.32        & 2.41       & 2.00 & 7-membered aromatic
carbon (e.g., azulene) \\
\hline
FGA3    & \textbf{0.19} & \textbf{0.17} & \textbf{0.15} & Aliphatic
fluorine, trifluoro group \\
FGR1    & \textbf{0.40} & \textbf{0.29} & \textbf{0.31} & Aromatic fluorine \\
\hline
HGA1    & \textbf{0.34} & \textbf{0.28} & \textbf{0.29} & Aliphatic
proton, CH \\
HGA2    & \textbf{0.27} & \textbf{0.21} & \textbf{0.24} & Aliphatic
proton, CH2 \\
HGA3    & \textbf{0.23} & \textbf{0.20} & \textbf{0.21} & Aliphatic
proton, CH3 \\
HGP1    & \textbf{0.36} & \textbf{0.31} & \textbf{0.33} & Polar hydrogen \\
HGP4    & \textbf{0.31} & \textbf{0.22} & \textbf{0.22} & Polar
hydrogen, neutral conjugated -NH2 group  \\
HGPAM1  & \textbf{0.44} & \textbf{0.38} & \textbf{0.37} & Polar
hydrogen, neutral dimethylamine group \\
HGPAM2  & \textbf{0.27} & \textbf{0.25} & \textbf{0.24} & Polar
hydrogen, neutral methylamine group \\
HGR51   & \textbf{0.38} & \textbf{0.34} & \textbf{0.39} & Nonpolar
hydrogen, neutral 5-membered planar ring carbon \\
HGR52   & \textbf{0.36} & \textbf{0.32} & \textbf{0.34} & Aldehyde
hydrogen, formamide hydrogen (RCOH);\\
HGR61   & \textbf{0.41} & \textbf{0.32} & \textbf{0.34} & Aromatic hydrogen \\
HGR62   & \textbf{0.40} & \textbf{0.31} & \textbf{0.33} & Nonpolar
hydrogen, neutral 6-membered planar ring \\
\hline
NG1T1   & \textbf{0.32} & \textbf{0.27} & \textbf{0.29} & Nitrogen in
cyano group \\
NG2D1   & 0.88        & 0.76       & 0.78 & Nitrogen in neutral imine \\
NG2R50  & \textbf{0.48} & \textbf{0.42} & \textbf{0.46} &
Double-bonded nitrogen in neutral 5-membered planar ring  \\
NG2R60  & 0.57        & \textbf{0.47} & \textbf{0.49} & Double-bonded
nitrogen in neutral 6-membered planar ring \\
NG2R62  & 0.61        & \textbf{0.48} & \textbf{0.49} & Double-bonded
nitrogen in 6-membered planar ring with heteroatoms  \\
NG321   & 1.28        & 1.21       & 1.18 & Neutral methylamine nitrogen \\
\hline
OG2D1   & \textbf{0.45} & \textbf{0.37} & \textbf{0.39} & Carbonyl oxygen  \\
OG2D4   & \textbf{0.47} & \textbf{0.37} & \textbf{0.37} & 6-membered
aromatic carbonyl oxygen  \\
OG2R50  & 0.58        & 0.50       & 0.57 & Oxygen in furan ring  \\
OG301   & 0.99        & 0.80       & 0.79 & Ether oxygen  \\
OG311   & 0.56        & \textbf{0.45} & \textbf{0.48} & Hydroxyl oxygen \\
OG3R60  & 0.76        & 0.62       & 0.63 & Oxygen in 6-membered
cyclic enol, ether, or ester \\
\bottomrule
\end{tabular}}

\end{table}

\end{document}